\documentclass[superscriptaddress, twocolumn, pra,nofootinbib]{revtex4-2}
\usepackage{graphicx}
\usepackage{calc}
\usepackage{amsmath}
\usepackage{amssymb}
\usepackage{color}
\usepackage[pdftex,dvipsnames,usenames]{xcolor}
\usepackage[colorlinks=true,urlcolor=blue,citecolor=blue,linkcolor=blue]{hyperref}
\usepackage{multirow}
\usepackage{array}

\usepackage[colorinlistoftodos,prependcaption,textsize=tiny]{todonotes}

\DeclareMathOperator{\Tr}{Tr}

\renewcommand{\Re}{\mbox{\rm Re}}

\DeclareMathOperator{\I}{I}

\begin{document}

\title{Numerical simulations of atmospheric quantum channels}

\author{M. Klen}
\affiliation{Bogolyubov Institute for Theoretical Physics, NAS of Ukraine, Vulytsya Metrologichna 14b, 03143 Kyiv, Ukraine}
\affiliation{Department of Theoretical and Mathematical Physics, Kyiv Academic University, Boulevard Vernadskogo  36, 03142  Kyiv, Ukraine}
\author{A. A. Semenov}
\affiliation{Bogolyubov Institute for Theoretical Physics, NAS of Ukraine, Vulytsya Metrologichna 14b, 03143 Kyiv, Ukraine}
\affiliation{Department of Theoretical and Mathematical Physics, Kyiv Academic University, Boulevard Vernadskogo  36, 03142  Kyiv, Ukraine}
\affiliation{Institute of Physics, NAS of Ukraine, Prospect Nauky 46, 03028 Kyiv, Ukraine}

\begin{abstract}
	Atmospheric turbulence is one of the lead disturbance factors for free-space quantum communication.
	The quantum states of light in such channels are affected by fluctuating losses characterized by the probability distribution of transmittance (PDT).
	We obtain the PDT for different horizontal links via numerical simulations of light transmission through the atmosphere.
	The results are compared with analytical models: the truncated log-normal distribution, the beam-wandering model, the elliptic-beam approximation, and the model based on the law of total probability.
	Their applicability is shown to be strongly dependent on the receiver aperture radius.
	We introduce an empirical model based on the Beta distribution, which is in good agreement with numerical simulations for a wide range of channel parameters.
	However, there are still scenarios where none of the above analytical models fits the numerically simulated data.
	The numerical simulation is then used to analyze the transmission of quadrature-squeezed light through free-space channels.
\end{abstract}

\maketitle


\section{Introduction}
\label{Sec:Intro}

    Quantum communication is a rapidly developing field with a wide range of applications.
	One of its main advantages is the ability to establish secure links between remote parties (for a review see Refs.~\cite{gisin02,Xu2020,Pirandola2020,Renner2023}).
	It can also be used in many other applications, such as quantum digital signature \cite{Gottesman01}, connecting quantum devices with quantum teleportation \cite{bennett93,Braunstein1998}, and entanglement swapping protocols \cite{Zukowski1993}, to name just a few.

    Optical radiation fields serve as natural carriers of quantum information for communication channels.
    Quantum light can be distributed through optical fibers or free space.
    The latter offers several practical advantages, including the ability to establish satellite-mediated global quantum communication, communication through hard-to-access regions, and communication with moving parties.
    Various implementations of free-space channels have been reported in the literature for horizontal (see, e.g., Refs.~\cite{ursin07,elser09,heim10,fedrizzi09,capraro12,yin12,ma12,peuntinger14,vasylyev17,Jin2019,Bulla2023}) and vertical (see, e.g., Refs.~\cite{nauerth13,wang13,bourgoin15}) links, including experiments involving satellites \cite{vallone15,dequal16,vallone16,carrasco-casado16,takenaka17,liao17,yin17,gunthner17,ren17,yin17b,Liao2018,Vasylyev2019,Ecker2021,Ecker2022}.


	Atmospheric turbulence is one of the major obstacles for free-space quantum communication.
	Its effect on the optical fields is described by methods \cite{Tatarskii,Fante1975,Fante1980} developed in classical atmospheric optics.
	The corresponding techniques differ depending on the degrees of freedom of light and the types of measurements involved in the protocol.
	For example, protocols dealing with optical angular momentum \cite{DAmbrosio2012,vallone2014,krenn2016} require a proper description of the spatial structure of the modes \cite{paterson05,pors11,Roux2013,Roux2015,Leonhard2015,Lavery2018,Sorelli2019}.
	Other types of protocols and experiments involving photocounting \cite{perina73,milonni04}, polarization analysis \cite{ursin07,fedrizzi09,semenov10,gumberidze16}, homodyne detection \cite{elser09,heim10,usenko12,semenov12,peuntinger14,heim14}, and so on, demand a proper description of the fluctuating losses in atmospheric channels.
	In this case, the light mode at the transmitter can be prepared, for example, as a quasi monochromatic mode in the form of a pulsed Gaussian beam.
	For discrete-variable protocols, using propagation of single photons in each direction and/or measurements with long detection times, fluctuations of losses (scintillation) can be neglected \cite{shapiro11,bonato09,lanning2021}.


   	In this paper we consider communication protocols with a fixed spatial structure of light modes at the transmitter.
   	As mentioned, in this case the quantum states of a quasi monochromatic light mode with a fluctuating shape at the receiver are affected by fluctuating losses.
   	They change the quantum states of the quasi monochromatic mode according to the input-output relation (cf. Refs.~\cite{semenov09,semenov12,vasylyev12} and Appendix~\ref{App:IOR}),
		\begin{align}\label{Eq:POutDist}
			P_\mathrm{out}\left(\alpha\right)=\int\limits_{0}^1 d\eta  \frac{1}{\eta}P_\mathrm{in}\left(\frac{\alpha}{\sqrt{\eta}}\right)\mathcal{P}(\eta).
		\end{align}
    Here $P_\mathrm{out}(\alpha)$ and $P_\mathrm{in}(\alpha)$ are the Glauber-Sudarshan $P$ functions \cite{glauber63c,sudarshan63} at the channel output and input, respectively, $\eta$ is the channel transmittance, and $\mathcal{P}(\eta)$ is the probability distribution of transmittance (PDT).
    Obviously, the PDT is the main characteristic of quantum channels.
    Equation~(\ref{Eq:POutDist}) has a similar form in both quantum and classical optics and thus the PDT can be obtained from purely classical considerations.


	The mentioned approach has been successfully applied to analyze quantum-state transfer through free-space channels (see Ref.~\cite{semenov2018} for a brief summary).
	For example, quadrature squeezing \cite{semenov12,vasylyev12}, Bell nonlocality \cite{semenov10,gumberidze16}, Gaussian entanglement \cite{bohmann16,hosseinidehaj15a}, and higher-order nonclassical and entanglement phenomena \cite{bohmann17a,hosseinidehaj15b} have been analyzed in the context of their transfer through atmospheric channels.
	Also the input-output relation (\ref{Eq:POutDist}) has been applied to analyze quantum communication protocols such as the decoy state protocol \cite{Wang2018,vasylyev18}, continuous-variable quantum key distribution protocols \cite{usenko12,guo2017,papanastasiou2018, derkach2020b,hosseinidehaj2019,ShiyuWang2018,chai2019, Derkach2021,pirandola2021,hosseinidehaj2021,pirandola2021b,pirandola2021c}, and continuous-variable quantum teleportation \cite{hofmann2019,zhang2017,villasenor2021}.

    A rigorous description of free-space quantum channels requires knowledge of the PDT.
    Random fluctuations of the refractive index due to atmospheric turbulence lead to random wandering of the beam centroid, fluctuations of the beam-spot shape, etc.
    These effects in turn lead to fluctuations in the intensity of the beam passing through the receiver aperture.
    The PDT characterizes the quantum channel associated with a quasimonochromatic mode.
    This characteristic can be obtained from purely classical considerations, corresponding to coherent states at the transmitter.

    Several analytical models of the PDT have been proposed.
	One of them focuses on the effect of beam wandering \cite{vasylyev12}.
	Furthermore, the elliptic-beam model \cite{vasylyev16} involves beam-spot distortions approximated by Gaussian elliptic beams with randomly oriented semiaxes.
	In particular, this model shows that beam-spot distortions can be effectively described by the truncated log-normal distribution considered earlier in Refs.~\cite{perina73,milonni04}.
	Assuming that beam wandering and broadening are statistically independent, the law of total probability can be employed to derive the PDT \cite{vasylyev18}.


    All analytical models have a common drawback: They are applicable for limited values of channel parameters such as turbulence strength, propagation distance, receiver aperture size, and beam parameters.
    It is usually difficult to determine which model to use for given values of the channel parameters.
    In this paper we address this problem by providing numerical simulations of the PDT for horizontal links and comparing the results with analytical models.
    For this purpose, we use the sparse-spectrum model \cite{Charnotskii2013a,Charnotskii2013b,Charnotskii2020} of the phase-screen method (see Refs.~\cite{Fleck1976,Frehlich2000,Lukin_book,Schmidt_book} and Refs.~\cite{Lushnikov2018,Lavery2018,Sorelli2019,Klug2023,Bachmann2023} for its recent applications).
    We also numerically check statistical properties of beams after passing through the atmosphere: Gaussianity of the distribution of the beam-centroid position, correlation between beam deflection and beam shape, etc.
    These properties are directly related to the applicability of analytical models.

    The phase-screen method cannot be considered a substitute for experimental investigations, since it involves approximations.
    Nevertheless, the sparse-spectrum model yields the statistics of phase disturbances that are in good agreement with the analytical expression (cf. Appendix~\ref{App:Verif}).
    Most importantly, the method is free of assumptions about the beam shape and its statistical characteristics made in the analytical models of the PDT.
    Thus, our results can be useful to determine their applicability.
    Numerically simulated data can also be used directly to analyze quantum light transmission in atmospheric channels for a variety of scenarios.
    Taking the quadrature squeezing as an instance, we demonstrate how numerical analysis can be employed to study the transmission of nonclassical phenomena through free-space channels.


    The rest of the paper is organized as follows.
	In Sec.~\ref{Sec:Preliminaries} we introduce preliminary information on the propagation of Gaussian beams through the turbulent atmosphere, the sparse-spectrum formulation of the phase-screen method, and analytical models.
	An empirical analytical model based on the Beta distribution, which shows good agreement with the numerically simulated data, is introduced in Sec.~\ref{Sec:BetaModel}.
	Numerical simulations of atmospheric channels and comparison of the obtained numerically-simulated data with analytical models are considered in Sec.~\ref{Sec:NumSim}.
	In Sec.~\ref{Sec:Stat} the statistical properties of beams after passing through the turbulent atmosphere are analyzed.
	An application of the discussed method to a description of the transmission of quadrature squeezing through free-space channels is considered in Sec.~\ref{Sec:Appl}.
	A summary and concluding remarks are given in Sec.~\ref{Sec:Concl}.
	In the Supplemental Material \cite{supplement} one can find numerically simulated data for different channels, an interactive visualization tool, and the corresponding Python 3 codes.


\section{Preliminaries}
\label{Sec:Preliminaries}

	In this section we briefly introduce preliminary results needed for our consideration.
	All of them are related to channel characterization by the PDT and are thus derived from classical considerations.
	First, we discuss the fluctuating channel transmittance $\eta$.
	Then we briefly review the sparse-spectrum formulation of the phase-screen method used for our numerical simulations.
	Finally, we review the existing analytical models of the PDT.


\subsection{Channel transmittance}

    The distribution of a light beam through the turbulent atmosphere along the $z$ axis can be described by the paraxial equation \cite{Fante1975} for the field amplitude $u(\mathbf{r};z)$,
		\begin{align}\label{Eq:Paraxial}
			2ik\frac{\partial u(\mathbf{r};z)}{\partial z}+\Delta_\mathbf{r}
			u(\mathbf{r};z)+2k^2\delta n(\mathbf{r},z) u(\mathbf{r};z)=0,
		\end{align}
	where $k$ is the wave number, $\delta n(\mathbf{r},z)$ is a fluctuating part of the index of refraction of air, and $\mathbf{r}=\begin{pmatrix} x & y\end{pmatrix}^T$ is the vector of transverse coordinates.
	The Gaussian beams are defined by the boundary conditions of this equation,
		\begin{align}\label{Eq:BoundaryConditions}
			u(\mathbf{r};0)=\sqrt{\frac{2}{\pi
					W_0^2}}\exp\Bigl[-\frac{\mathbf{r}^2}{W_0^2}{-}\frac{ik
			}{2F_0}\mathbf{r}^2\Bigr].
		\end{align}
	Here $W_0$ and $F_0$ are the beam-spot radius and the wave-front radius at the transmitter, respectively.
	For our purposes we will use two types of initial conditions.
	The first corresponds to the case of $F_0=+\infty$.
	In the absence of atmosphere, this beam is collimated i.e. it is minimally divergent, for a propagation distance much shorter than the Rayleigh length, $z\ll z_\mathrm{R}=kW_0^2/2$.
	The second corresponds to $F_0=z_\mathrm{ap}$, where $z_\mathrm{ap}$ is the channel length.
	Such a beam is considered to be geometrically focused (cf. Ref.~\cite{Andrews_book}) because in the absence of atmosphere it has a minimum beam spot radius in the plane of the receiver aperture, $z=z_\mathrm{ap}$, given $W_0$ and $z_\mathrm{ap}$.

    Random fluctuations of the index of refraction in the case of isotropic homogeneous turbulence are described by the correlation function, the Markovian approximation for which reads
        \begin{align}\label{Eq:Correlator_n_n}
            \langle\delta n&(\mathbf{r}_1;z_1)\delta n(\mathbf{r}_2;z_2)\rangle\nonumber\\
            &=\int_{\mathbb{R}^2}d^2\boldsymbol{\kappa}\Phi_n(\boldsymbol{\kappa};\kappa_z{=}0)e^{i\boldsymbol{\kappa}\cdot(\mathbf{r}_1-\mathbf{r}_2)}\delta(z_1-z_2),
        \end{align}
    where $\Phi_{n}(\boldsymbol{\kappa};\kappa_z)$ is the turbulence spectrum.
    We will use it in the modified von K\'arm\'an--Tatarskii form given by
        \begin{align}\label{Eq:Karman}
            \Phi_n(\boldsymbol{\kappa};\kappa_z)=\frac{0.033 C_n^2\exp\left[-\left(\frac{\kappa\ell_0}{2\pi}\right)^2\right]}{\left(\kappa^2+L_0^{-2}\right)^{11/6}}.
        \end{align}
    Here $C_n^2$ is the index-of-refraction structure constant characterizing the local strength of the turbulence, $\ell_0$ and $L_0$ are inner and outer scales of turbulence corresponding to minimum and maximum sizes of the turbulence eddies, respectively, and $\kappa=\sqrt{\boldsymbol{\kappa}^2+\kappa_z^2}$.

    The beam intensity is defined as $I(\mathbf{r},z)=|u(\mathbf{r};z)|^2$.
	As discussed in Refs.~\cite{vasylyev12,vasylyev16,vasylyev18} (see also Appendix~\ref{App:IOR}), the transmittance is given by integration of the field intensity over the aperture opening $\mathcal{A}$ in the receiver aperture plane $z=z_\mathrm{ap}$,
		\begin{align}\label{Eq:Efficiency}
			\eta=\int_\mathcal{A}d^2\mathbf{r}I(\mathbf{r};z_\mathrm{ap}).
		\end{align}
	Fluctuations of $\delta n(\mathbf{r};z)$ lead to fluctuations of $I(\mathbf{r};z)$ and consequently to randomization of $\eta$.
	Moreover, this consideration does not take into account deterministic losses caused by absorption and scattering, which can be easily included in the final results.

    Next, we consider the second- and fourth-order correlation functions given by
        \begin{align}\label{Eq:Gamma2}
            \Gamma_2(\mathbf{r};z)=\left\langle I(\mathbf{r};z)\right\rangle,
        \end{align}
    and
        \begin{align}\label{Eq:Gamma4}
            \Gamma_4(\mathbf{r}_1,\mathbf{r}_2;z)=\left\langle I(\mathbf{r}_1;z)I(\mathbf{r}_2;z)\right\rangle,
        \end{align}
    respectively.
    These functions can be used to obtain various statistical properties of the light beam passing through the turbulent atmosphere.
    For instance, the first two moments of the channel transmittance can be evaluated as
        \begin{align}\label{Eq:Eta}
            \langle\eta\rangle=\int_\mathcal{A}d^2\mathbf{r}\Gamma_2(\mathbf{r};z)
        \end{align}
    and
        \begin{align}\label{Eq:Eta2}
            \langle\eta^2\rangle=\int_\mathcal{A}d^2\mathbf{r}_1\int_\mathcal{A}d^2\mathbf{r}_2\Gamma_4(\mathbf{r}_1,\mathbf{r}_2;z).
        \end{align}
    Other applications of these functions needed for analytical models of the PDT will be discussed in Sec.~\ref{Sec:AnalyticModels}.
	In this context, it is also important to mention the recent results in Refs.~\cite{Klug2023,Bachmann2023} about spatial modes that are different from Gaussian beams and which exhibit better transmittance for particular realizations of turbulent channels.


\subsection{Sparse-spectrum model for phase-screen method}
\label{Sec:PS}

	According to Eq. (\ref{Eq:Efficiency}), in order to sample the channel transmittance $\eta$, one must first sample the field amplitude $u(\mathbf{r};z)$.
	This can be accomplished using the sparse-spectrum model for the phase-screen method \cite{Charnotskii2013a,Charnotskii2013b,Charnotskii2020}.
	Its verification technique is discussed in Appendix~\ref{App:Verif}.
	The method relies on the fact that given the field amplitude $u(\mathbf{r}, z_{m-1})$ at the point $z=z_{m-1}$, one can find an approximate solution to the paraxial equation (\ref{Eq:Paraxial}) at the point $z=z_m>z_{m-1}$  up to the second order of $l=z_m-z_{m-1}$ (cf. Ref.~\cite{Fleck1976}),
		\begin{align}\label{Eq:splitoperator}
			u(\mathbf{r};z_m)
			\approx e^{-({il}/{4k})\Delta_\mathbf{r}} e^{-i\phi(\mathbf{r})} e^{-({il}/{4k})\Delta_\mathbf{r}} u(\mathbf{r};z_{m-1}).
		\end{align}
	Here
		\begin{align}\label{Eq:Phase}
			\phi(\mathbf{r})= k\int\limits_{z_{m-1}}^{z_m} d z\, \delta n(\mathbf{r},z)
		\end{align}
	is referred to as the phase screen.
	As follows from Eq.~(\ref{Eq:Correlator_n_n}), the correlation function of the latter is
		\begin{align}\label{Eq:Correlator_phi}
			\langle \phi(\mathbf{r}_1) 	\phi(\mathbf{r}_2)\rangle=\int_{\mathbb{R}^2}d^2\boldsymbol{\kappa}\Phi_\phi(\boldsymbol{\kappa})e^{i\boldsymbol{\kappa}\cdot(\mathbf{r}_1-\mathbf{r}_2)},
		\end{align}
	where
		\begin{align}
			\Phi_\phi( \boldsymbol{\kappa}) = 2\pi l k^2 \Phi_n(\boldsymbol{\kappa};0)
		\end{align}
	is the power spectral density of the phase screens.
	Equation (\ref{Eq:splitoperator}) implies that propagation between the points $z_{m-1}$ and $z_{m}$ is reduced to vacuum propagation up to the midpoint of this interval, followed by the phase increment by $\phi(\mathbf{r})$ and finally vacuum propagation again.

    The phase-screen method can be outlined as follows.
    First, the propagation distance is divided into $M$ slabs $[z_{m-1},z_m]$, where $m=1, \ldots, M$, $z_0=0$, and $z_M=z_\mathrm{ap}$.
    Then the phase screens are sampled at the midpoints of these slabs.
    Finally, the vacuum propagation of the field amplitude $u(\mathbf{r}, z)$ is simulated successively between these points and the $\mathbf{r}$-dependent phase is incremented on them.

	The vacuum propagation can be simulated using the Fourier transform of the field amplitude.
	In the sparse-spectrum model \cite{Charnotskii2013a, Charnotskii2013b, Charnotskii2020}, the phase screens are expanded into $N$ harmonics such that
	   \begin{align}
		   	\phi(\mathbf{r})\approx \Re\sum_{n=1}^N a_n e^{i\boldsymbol{\kappa}_n\cdot\mathbf{r}}.
		\end{align}
	The sampled values of $\phi(\mathbf{r})$ are obtained via sampling the expansion coefficients $a_n$ and the random vectors $\boldsymbol{\kappa}_n$.
	Their statistics should satisfy Eq.~(\ref{Eq:Correlator_phi}).
	This can be achieved if two conditions are fulfilled.
	First, the real and imaginary parts of the coefficients $a_n$ are normally distributed with
	   \begin{align}
		   	\langle a_n\rangle = 0, \quad \langle a_n a_m\rangle = 0,  \quad \langle a_n a^*_m\rangle = s_{n}\delta_{nm}.
	   \end{align}
	Second, the values of $s_n$ and the random vectors $\boldsymbol{\kappa}_n$ are chosen such that
	   \begin{align}\label{Eq:PSMCond}
		   	\sum_{n=1}^N s_n p_n(\boldsymbol{\kappa}) = 2 \Phi_\phi(\boldsymbol{\kappa}),
	   \end{align}
	where $p_n(\boldsymbol{\kappa}_n)$ are the probability distributions for the random vectors $\boldsymbol{\kappa}_n$.

	Let us split the domain of $\boldsymbol{\kappa}$ into $N$ rings, $K_{n-1}<\kappa<K_n$, where
	  \begin{align}
	   	K_n =K_\mathrm{min}\exp\left[\frac{n}{N}\ln\left(\frac{K_\mathrm{max}}{K_\mathrm{min}}\right)\right],
	  \end{align}
	$n=0, \ldots, N$, and $K_\mathrm{min}$ and $K_\mathrm{max}$ are the inner and outer bounds of the spectral domain, respectively.
	We define $p_n(\boldsymbol{\kappa}_n)$ such that it has zero values outside the corresponding ring.
	In this case, Eq.~(\ref{Eq:PSMCond}) is reduced to the form
	  \begin{align}\label{Eq:PSMCond1}
	   	s_n p_n(\boldsymbol{\kappa}_n) = 2 \Phi_\phi(\boldsymbol{\kappa}_n).
	  \end{align}
	Integrating this expression inside the ring, we get
	  \begin{align}
	   	s_n=2\pi \int\limits_{K_{n-1}}^{K_n} d\kappa\kappa\Phi_\phi(\kappa).
	  \end{align}
	The probability distributions for the vector $\boldsymbol{\kappa}_n$ are given by $p_n(\boldsymbol{\kappa}_n)=2s_n^{-1}\Phi_\phi(\boldsymbol{\kappa}_n)$.
	We approximate it by sampling the absolute values of $\boldsymbol{\kappa}_n$ as
	  \begin{align}
	   	\kappa_n =\sqrt{K_{n-1}^2 + \xi_n \left(K_n^2 - K_{n-1}^2\right)},
	  \end{align}
	where $\xi_n\in[0,1]$ is the uniformly distributed random variable, and by independently sampling the polar angle of $\boldsymbol{\kappa}_n$ as a uniformly distributed random variable in the domain $[0,2\pi]$.


\subsection{Analytical and semi-analytical models}
\label{Sec:AnalyticModels}


\subsubsection{Truncated log-normal model}

    The log-normal distribution is a widely used model that describes light passing through atmospheric channels \cite{perina73,milonni04,capraro12}.
	In the domain of its applicability, this distribution should be truncated, since the channel efficiency cannot exceed the value of one.
	The normalized truncated log-normal PDT is given by
		\begin{align}\label{Eq:PDT_LN}
			\mathcal{P}(\eta;\mu,\sigma)
			=\frac{1}{\mathcal{F}(1)}\frac{1}{\sqrt{2\pi}\eta\sigma}\exp\Bigl[-\frac{(\ln \eta+\mu )^2}{2\sigma^2}\Bigr],
		\end{align}
	where $\mathcal{F}(x)$ is the cumulative distribution function of the log-normal distribution.
	The parameters $\mu$ and $\sigma$ are related to the first and second moments of the transmittance [cf. Eqs.~(\ref{Eq:Eta}) and (\ref{Eq:Eta2})], respectively, as
		\begin{align}\label{Eq:Mu}
			\mu=\mu\left(\langle\eta\rangle,\langle\eta\rangle^2\right)\approx-\ln\left[\frac{\langle\eta\rangle^2}{\sqrt{\langle\eta^2\rangle}}\right],
		\end{align}
		\begin{align}\label{Eq:Sigma}
			\sigma^2=\sigma^2\left(\langle\eta\rangle,\langle\eta\rangle^2\right)\approx\ln\left[\frac{\langle\eta^2\rangle}{\langle\eta\rangle^2}\right].
		\end{align}
	Thus, the truncated log-normal model is completely characterized by two moments of the transmittance, i.e., by the information contained in the field correlation functions (\ref{Eq:Gamma2}) and (\ref{Eq:Gamma4}).


\subsubsection{Beam-wandering model}

	Here we briefly sketch the beam-wandering model considered in Ref.~\cite{vasylyev12}.
	In this model it is assumed that the beam intensity at the receiver-aperture plane has a Gaussian shape, but $W_0$ is replaced with the short-term beam-spot radius $W_{\mathrm{ST}}$,
		\begin{align}\label{Eq:ST}
			W_{\mathrm{ST}}^2=W_{\mathrm{LT}}^2- 4\sigma_{\mathrm{bw}}^2.
		\end{align}
	Here $W_{\mathrm{LT}}$, defined as
		\begin{align}\label{Eq:LT}
			W_{\mathrm{LT}}^2=4\int_{\mathbb{R}^2}d\mathbf{r} x^2\Gamma_2(\mathbf{r};z),
		\end{align}
	is the long-term beam-spot radius and
		\begin{align}\label{Eq:SigmaBW}
			\sigma_{\mathrm{bw}}^2=\int_{\mathbb{R}^4}d\mathbf{r}_1d\mathbf{r}_2 x_1 x_2\Gamma_4(\mathbf{r}_1,\mathbf{r}_2;z)
		\end{align}
	is the variance of a beam-centroid coordinate.
	Assuming that the beam-centroid position is normally distributed, an analytical form for the PDT is derived in Ref.~\cite{vasylyev12}, (see also Appendix~\ref{App:BW}).
	The basic assumption behind this model is that atmospheric turbulence results mostly in beam wandering.


\subsubsection{Elliptic-beam model}

    The main idea behind the elliptic-beam model \cite{vasylyev16} is to approximate the beam intensity by the elliptic Gaussian form
        \begin{align}\label{Eq:EllipticIntens}
            I(\mathbf{r};z_\mathrm{ap})=\frac{2}{\pi\sqrt{\det\mathbf{S}}}\exp\Bigl[-2({\mathbf{r}}{-}{
            \mathbf{r}}_0)^{\rm
            T}{\mathbf{S}}^{-1}({\bf r}{-}{\bf r}_0)\Bigr].
        \end{align}
    Here
        \begin{align}\label{Eq:BeamCentroid}
            \mathbf{r}_0=\int_{\mathbb{R}^2}d^2\mathbf{r}\,\mathbf{r}\,
            I(\mathbf{r};z_\mathrm{ap})
        \end{align}
    and
        \begin{align}\label{Eq:MatrixS_Definition}
            \mathbf{S}=4\int_{\mathbb{R}^2}d^2\mathbf{r}
            \left[(\mathbf{r}-\mathbf{r}_0)(\mathbf{r}-\mathbf{r}_0)^\mathrm{T}
            \right ]
            I(\mathbf{r};z_\mathrm{ap})
        \end{align}
    are the beam-centroid position and the spot-shape matrix, respectively.
    As it has been discussed in Ref.~\cite{vasylyev16}, all parameters of the elliptic-beam model can be approximately related to the field correlation functions (\ref{Eq:Gamma2}) and (\ref{Eq:Gamma4}) (see also Appendix~\ref{App:EB}).

    In this paper, along with the analytical approximation considered in Ref.~\cite{vasylyev16}, we will use a semianalytical technique to obtain the elliptic-beam PDT.
    This method is free from assumptions about the statistical characteristics of the elliptic-beam model.
    In particular, this technique involves the following steps: (i) The values of the field intensity $I(\mathbf{r};z_\mathrm{ap})$ are sampled using the sparse-spectrum model for the phase-screen method, (ii) the sampled values of $\mathbf{r}_0$ and $\mathbf{S}$ are obtained using Eqs. ~(\ref{Eq:BeamCentroid}) and (\ref{Eq:MatrixS_Definition}), (iii) these values are successively substituted into Eqs.~(\ref{Eq:EllipticIntens}) and (\ref{Eq:Efficiency}) to obtain the sampled values of the transmittance $\eta$, and (iv) the elliptic-beam PDT is reconstructed from the sampled values of $\eta$.
    This semi-analytical approach is not considered a practical tool for obtaining the PDT.
    Rather, it is used solely to assess the general applicability of the elliptic-beam model, without relying on analytical approximations of the model parameters.


\subsubsection{The model based on the law of total probability}
\label{Sec:LTP}

    As discussed in Ref.~\cite{vasylyev16}, the beam-spot distortion leads to the log-normal shape of the PDT.
 	At the same time, beam wandering leads to the log-negative Weibull distribution \cite{vasylyev12}.
 	The model based on the law of total probability \cite{vasylyev18} employs these observations and combines them into a single PDT that accounts for two effects simultaneously.
 	The corresponding function is given by
		 \begin{align}\label{Eq:PDT_Bayesian}
		 	\mathcal{P}(\eta)=\int_{\mathbb{R}^2} d^2 \boldsymbol{r}_0 \mathcal{P}(\eta|\boldsymbol{r}_0)\rho(\boldsymbol{r}_0).
		 \end{align}
	Here $\rho(\boldsymbol{r}_0)$ is the Gaussian probability distribution for a beam-centroid coordinate with variance (\ref{Eq:SigmaBW}).
    The function
        \begin{align}
            \mathcal{P}(\eta|\boldsymbol{r}_0)=\mathcal{P}(\eta;\mu_{r_0},\sigma_{r_0})
        \end{align}
    is the truncated log-normal PDT [cf. Eq.~(\ref{Eq:PDT_LN})], where $\mu_{r_0}=\mu\left(\langle\eta\rangle_{r_0},\langle\eta^2\rangle_{r_0}\right)$ and $\sigma_{r_0}=\sigma\left(\langle\eta\rangle_{r_0},\langle\eta^2\rangle_{r_0}\right)$ are obtained from Eqs.~(\ref{Eq:Mu}) and (\ref{Eq:Sigma}), respectively.
    The conditional moments $\langle\eta\rangle_{r_0}$ and $\langle\eta^2\rangle_{r_0}$ are defined similarly to (\ref{Eq:Eta}) and (\ref{Eq:Eta2}), respectively.
    However, in this case the field correlation functions are replaced by $\Gamma_2^{(c)}(\mathbf{r};z) =\left\langle I^{(c)}(\mathbf{r}; z)\right\rangle$ and $\Gamma_4^{(c)}(\mathbf{r}_1,\mathbf{r}_2;z)=\left\langle I^{(c)}(\mathbf{r}_1;z)I^{(c)}(\mathbf{r}_2;z)\right\rangle$, where
    \begin{align}\label{Eq:Ic}
    	I^{(c)}(\boldsymbol{r};z)=I(\boldsymbol{r}+\boldsymbol{r}_0;z)
    \end{align}
    is the beam intensity function with the origin of the coordinate system placed at the position $\boldsymbol{r}_0$ of the beam centroid.

	There are two ways to obtain the conditional moments $\langle\eta\rangle_{r_0}$ and $\langle\eta^2\rangle_{r_0}$.
	The first requires direct integration of the field correlation functions $\Gamma_2^{(c)}(\mathbf{r};z)$ and $\Gamma_4^{(c)}(\mathbf{r}_1,\mathbf{r}_2;z)$.
	This leads to an involved integration of the analytic expressions for these functions, which in turn are often expressed in terms of multiple integrals.
	However, given the numerically simulated data, these moments can be evaluated with standard numerical integration.
	The PDT obtained with such a semianalytical technique can be used for a general analysis of the model applicability.
	The second possibility is the analytical approximation based on the assumption of weak beam wandering (see Ref.~\cite{vasylyev18} and Appendix~\ref{App:TPL}).
	An analysis of the corresponding PDT will give us information about the suitability of this approximation.


\section{Beta-distribution model}
\label{Sec:BetaModel}

	In this section we introduce an empirical model based on the Beta distribution \cite{johnson_book}.
	The choice of such an ansatz is motivated by several reasons.
	First, this distribution is naturally defined in the domain $\eta\in[0,1]$.
	Then, it can be parametrized by the two moments $\langle\eta\rangle$ and $\langle\eta^2\rangle$.
	Finally, this distribution has a highly variable shape depending on the parameter values.
	We will show that the Beta-distribution model fits the results of numerical simulations better than other analytical models for a wide range of channel parameters.

    The PDT for the Beta-distribution model reads
        \begin{align}\label{Eq:pdt_beta}
            \mathcal{P}(\eta; a, b) = \frac{1}{B(a, b)} \eta^{a-1} (1-\eta)^{b-1}.
        \end{align}
    Here $B(a, b)$ is the Beta function.
    The parameters $a$ and $b$ are given by
        \begin{align}\label{Eq:pdt_beta_ab}
            &a = a\left(\langle\eta\rangle,\langle\eta^2\rangle\right)= \frac{\langle\eta\rangle - \langle\eta^2\rangle}{ \langle\eta^2\rangle- \langle\eta\rangle^2}\langle \eta \rangle, \\
            &b =b\left(\langle\eta\rangle,\langle\eta^2\rangle\right)=  a\left(\langle\eta\rangle,\langle\eta^2\rangle\right) \left( \frac{1}{\langle\eta\rangle} - 1 \right).
        \end{align}
    These expressions parametrize the PDT in terms of $\langle \eta\rangle$ and $\langle \eta^2\rangle$.

   	In many practical scenarios, the Beta PDT can be considered as a reasonable alternative to the truncated log-normal distribution.
    This implies that one can reformulate the model based on the law of total probability such that $\mathcal{P}(\eta|\boldsymbol{r}_0)$ in Eq.~(\ref{Eq:PDT_Bayesian}) is replaced with the Beta PDT as
    \begin{align}
    \mathcal{P}(\eta|\boldsymbol{r}_0)=\mathcal{P}(\eta;a_{r_0},b_{r_0}),
    \end{align}
    where $a_{r_0}=a\left(\langle\eta\rangle_{r_0},\langle\eta^2\rangle_{r_0}\right)$ and $b_{r_0}=b\left(\langle\eta\rangle_{r_0},\langle\eta^2\rangle_{r_0}\right)$.
    In this case, the conditional moments $\langle\eta\rangle_{r_0}$ and $\langle\eta^2\rangle_{r_0}$ are defined in the same way as discussed in Sec.~\ref{Sec:LTP}.


\section{Numerical simulations}
\label{Sec:NumSim}

	In this section we will simulate the PDTs for various horizontal channels and compare the results with the analytical models considered in Secs.~\ref{Sec:AnalyticModels} and ~\ref{Sec:BetaModel}.
	We will focus on three channels classified by the value of the Rytov parameter $\sigma_R^2=1.23C_n^2k^{7/6}z_\mathrm{ap}^{11/6}$ (see Ref.~\cite{Andrews_book}).
	This parameter characterizes the strength of scintillation and the overall impact of turbulence on the entire channel, taking into account both the local turbulence strength characterized by $C_n^2$ and the channel length $z_\mathrm{ap}$.
	We will consider three scenarios: weak impact of turbulence with $\sigma_\mathrm{R}^2<1$, moderate  impact of turbulence with $\sigma_\mathrm{R}^2=1, \ldots, 10$, and strong  impact of turbulence with $\sigma_\mathrm{R}^2>10$, as shown in Table~\ref{Tab:Regimes}.
	The numerically simulated data, an interactive tool for their visualization, and the corresponding code can be accessed in the Supplemental Material \cite{supplement}.
	It should be noted that our simulations do not include deterministic losses caused by absorption and scattering.
	For practical applications of our results, these losses should be additionally taken into account.
	In a typical scenario, they can be considered on the order of 0.1~dB/km (see Ref.~\cite{scarani09}).

        \begin{table}[h!]
    	\caption{\label{Tab:Regimes} Parameter values for channels with weak, moderate, and strong  impact of turbulence considered in this paper.
    	}
    	\renewcommand{\arraystretch}{1.6}
    	\setlength{\tabcolsep}{4pt}
    	\begin{tabular}{>{\raggedright}p{11em}ccc}
    		\hline
    		\hline
    		Parameter & Weak & Moderate & Strong\\[1ex]
    		\hline
    		Rytov parameter $\sigma_\mathrm{R}^2$ & 0.2 & 1.5 & 33.3 \\
    		structure constant $C_n^2$~(m$^{-2/3}$) & $5{\times}10^{-15}$ & $1.5{\times}10^{-14}$ &  $6{\times}10^{-16}$ \\
    		outer scale $L_0$~(m) & $80$ & $80$ & $80$ \\
    		inner scale $\ell_0$~(m) & $10^{-3}$ & $10^{-3}$ & $10^{-3}$ \\
    		channel length $z_\mathrm{ap}$~(km) & 1 & 1.6 & 50 \\
    		beam-spot radius at the transmitter, $W_0$~(cm) & 2 & 2 & 6 \\
    		wavelength $\lambda=2\pi/k$~(nm) & 809 & 809 & 808 \\
    		\hline
    		\hline
    	\end{tabular}
    \end{table}

	In order to quantify the differences between the analytical or semianalytical and numerical PDTs, we will use the Kolmogorov-Smirnov (KS) statistic (see, e.g., Ref.~\cite{Berger2014}), given by
		\begin{align}
	 		D_M = \sup_\eta \left| F_M(\eta) - F(\eta) \right|.
	 	\end{align}
	Here $F(\eta)=\int_0^{\eta} d\eta^\prime \mathcal{P}(\eta^\prime)$ is the cumulative probability distribution of transmittance corresponding to the analytical model, $F_M(\eta)=M^{-1}\sum_{i=1}^M \theta(\eta-\eta_i)$ is the empirical probability distribution obtained from $M$ sampled transmittances $\eta_i$, and $\theta(\eta)$ is the Heaviside step function.
    If an analytical or semi-analytical model presumes samples of transmittance, $F(\eta)$ is replaced by the empirical probability distribution obtained from the corresponding sampled values.
    The cases $D_M = 0$ and $D_M = 1$ correspond to complete equivalence and maximum discrepancy, respectively, between the numerically simulated data and analytical models.

	We simulate data for three channels with the technique described in Sec.~\ref{Sec:PS}.
	For the channels with weak and moderate  impact of turbulence, we use a spatial grid with 512 points along one axis. For the channel with strong  impact of turbulence, we use a spatial grid with 4096 points.
	The spatial grid steps are 0.3, 0.4, and 1~mm for the channels with weak, moderate, and strong  impact of turbulence, respectively.
	The number of spectral rings is $N=1024$.
	The inner and outer bounds of the spectrum are $K_\mathrm{min}=1/15 L_0$ and $K_\mathrm{max}=2/\ell_0$, respectively.
	The number of phase screens is chosen from the condition that the Rytov parameter for the interscreen distance does not exceed $0.1$ (cf.~Refs.~\cite{Schmidt_book,Martin1988}).
    It is equal to 10 for the channels with weak and moderate  impact of turbulence and 30 for the channel with strong  impact of turbulence.
	This implies that the values of the Rytov parameter for the interscreen distances in these cases are $3\times10^{-3}$, $2.2\times10^{-2}$, and $6.5\times10^{-2}$, respectively.
	Small values of this parameter enable proper simulation of atmospheric channels.
	The number of samples $M=10^5$ for all channels.


\subsection{Channel with weak impact of turbulence}
\label{Sec:Weak}

 	In this section we consider a 1 km channel with weak local turbulence, as detailed in the second column of Table~\ref{Tab:Regimes}.
 	These parameter values result in a weak  impact of turbulence over the entire channel.
 	For $F_0=+\infty$, the Rayleigh length is $z_\mathrm{R}=kW_0^2/2\approx1.553~\mathrm{km}$, which is greater than the channel length $z_\mathrm{ap}{=}1~\mathrm{km}$.
 	Therefore, in the absence of the atmosphere, this beam can be approximately considered collimated.
 	When $F_0=z_\mathrm{ap}$, the beam-spot radius at the receiver aperture plane is minimized in the absence of the atmosphere for a given value of $W_0$ and $z_\mathrm{ap}$.
 	Such a beam can be considered geometrically focused, as discussed in Ref.~\cite{Andrews_book}.

\begin{figure}[ht!]
            \includegraphics[width=1\linewidth]{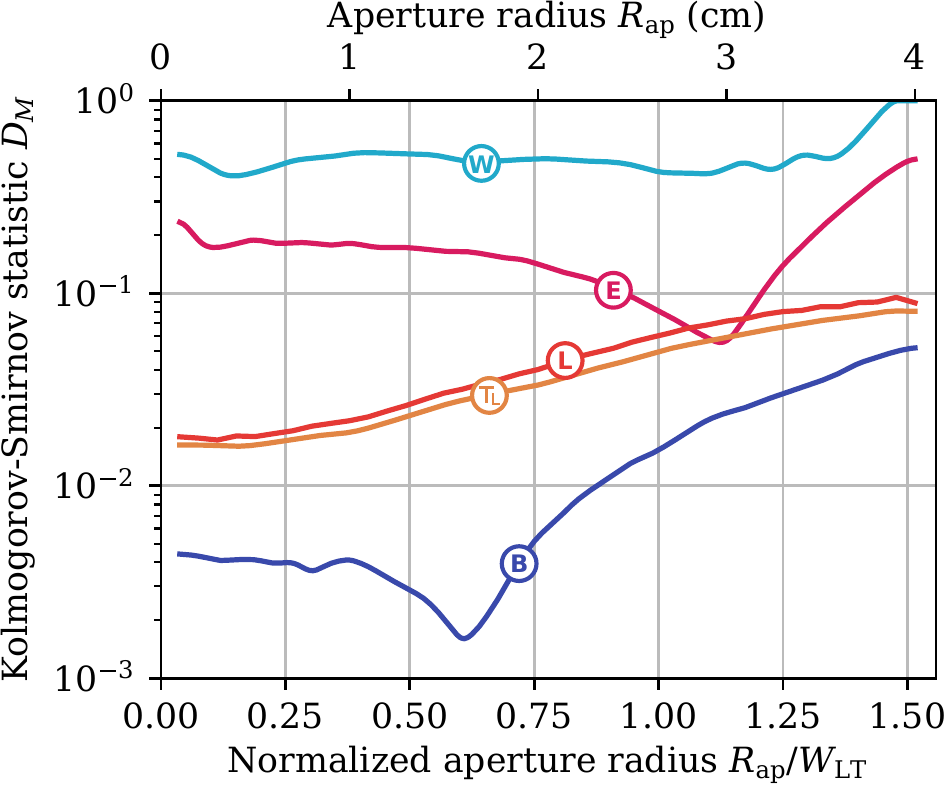}
			\caption{\label{Fig:weak_c_ks} The KS statistics, quantifying the differences between the numerically-simulated and analytical PDTs, shown for the beam-wandering model (W), the elliptic-beam model (E), the truncated log-normal model (L), the model based on the law of total probability with the approximation of weak beam wandering (T$_\textrm{L}$), and the Beta-distribution model (B) for the channel with weak  impact of turbulence and $F_0=+\infty$.
			}
		\end{figure}

\begin{figure}[h!]
            \includegraphics[width=1\linewidth]{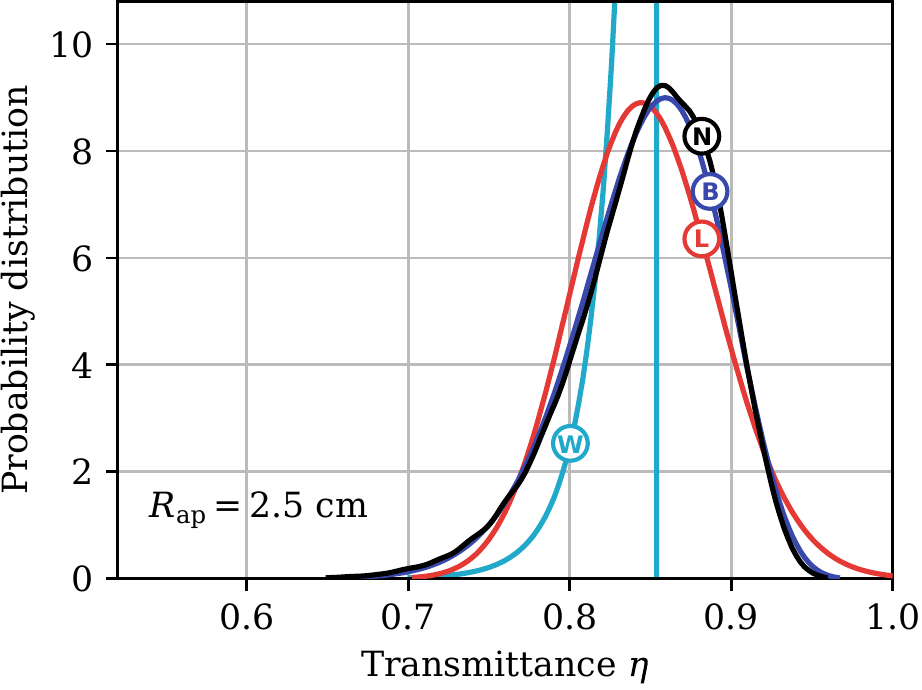}
			\caption{\label{Fig:weak_c_pdt} Numerically simulated (N), Beta-distribution (B), beam-wandering (W), and truncated log-normal (L) PDTs shown for the channel with weak impact of turbulence and $F_0=+\infty$.}
		\end{figure}

	First, consider the collimated beam.
	The dependence of the KS statistic on the aperture radius is shown in Fig.~\ref{Fig:weak_c_ks}.
	Here and in the following, the aperture radius is normalized by $W_{\mathrm{LT}}$ obtained according to Eq.~(\ref{Eq:LT}) from numerically simulated $\Gamma_2(\mathbf{r};z_{\mathrm{ap}})$.
	Evidently, the beam-wandering and elliptic-beam models show the worst agreement with the numerically simulated data.
	This result is counterintuitive since the beam-spot distortion is expected to be small for weak  impact of turbulence.
	However, our simulations show that the contribution of this effect is still significant compared to beam wandering.
	The Beta-distribution model shows the best agreement with the numerically simulated data for the considered channel.
	This model has a simple analytical form [cf. Eq.~(\ref{Eq:pdt_beta})] and can be easily used for analytical considerations of various nonclassical effects for quantum light distributed in such channels.
	An example of the numerically simulated PDT and different analytical PDTs can be seen in Fig.~\ref{Fig:weak_c_pdt}.
	This example clearly demonstrates that the Beta-distribution model has the best agreement with the numerically simulated PDT in this case.
	The truncated log-normal and beam-wandering models show much worse agreement.

\begin{figure}[h!]
			\includegraphics[width=1\linewidth]{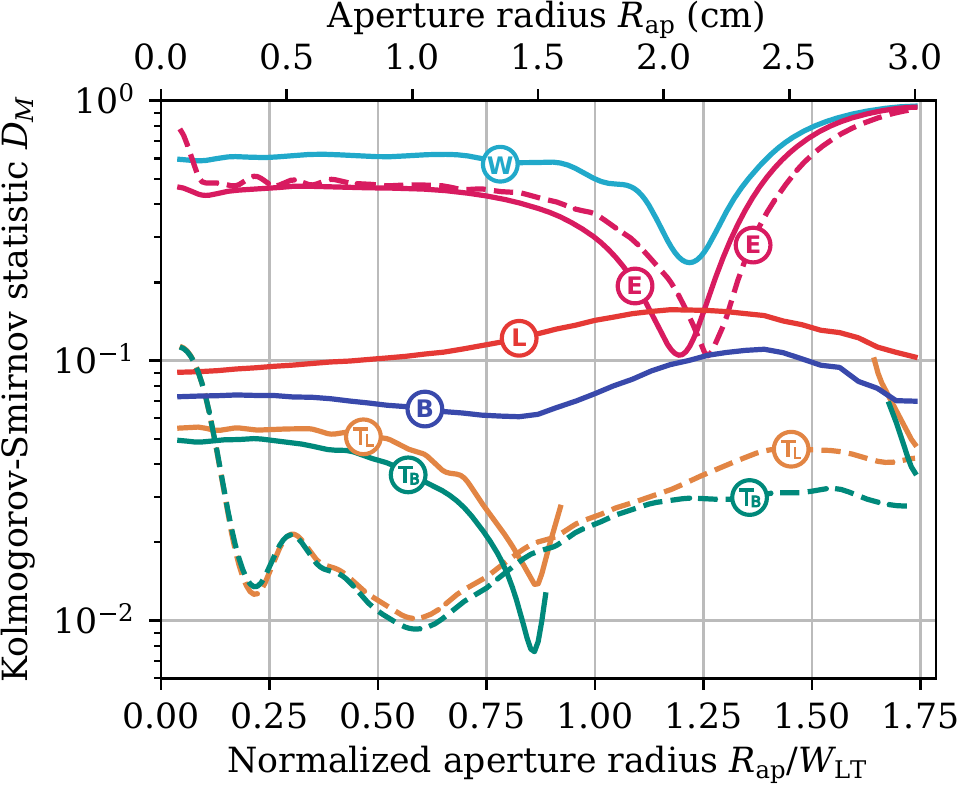}
			\caption{\label{Fig:weak_f_ks}  The KS statistics, quantifying the differences between the numerically-simulated and analytical PDTs, shown for the beam-wandering model (W), the elliptic-beam model (E), the truncated log-normal model (L), the model based on the law of total probability with the truncated log-normal (T$_\mathrm{L}$) and Beta  (T$_\textrm{B}$) distributions, and the Beta-distribution model (B) for the channel with weak impact of turbulence and  $F_0=z_\mathrm{ap}$.
			Solid and dashed lines correspond to analytical and semianalytical models, respectively.}
		\end{figure}

	Second, we consider the geometrically focused beam.
	The corresponding KS statistic as a function of aperture radius is shown in Fig.~\ref{Fig:weak_f_ks}.
	The best result for small values of the aperture radius is demonstrated by the total probability model based on the Beta distribution.
	For the remaining domain, this approximation no longer works.
	However, the model based on the law of total probability without the assumption of weak beam wandering still shows the best agreement with the numerically simulated data, as can be seen from the KS statistic for the corresponding semianalytical model.
	It is important to note that the estimation of the PDT for such semianalytical models is time consuming.
	For this reason, it is performed only with $5\times10^3$ samples, leading to a significant increase in the statistical error for the KS statistics.
	The elliptic-beam model, while showing good agreement with the numerically simulated data in the global minimum domain for the KS statistic, has a small discrepancy between the semianalytical and analytical models.
	The best choice among the analytical models for large values of the aperture radius is given by the Beta-distribution model.

		\begin{figure}[h!]
			\includegraphics[width=1\linewidth]{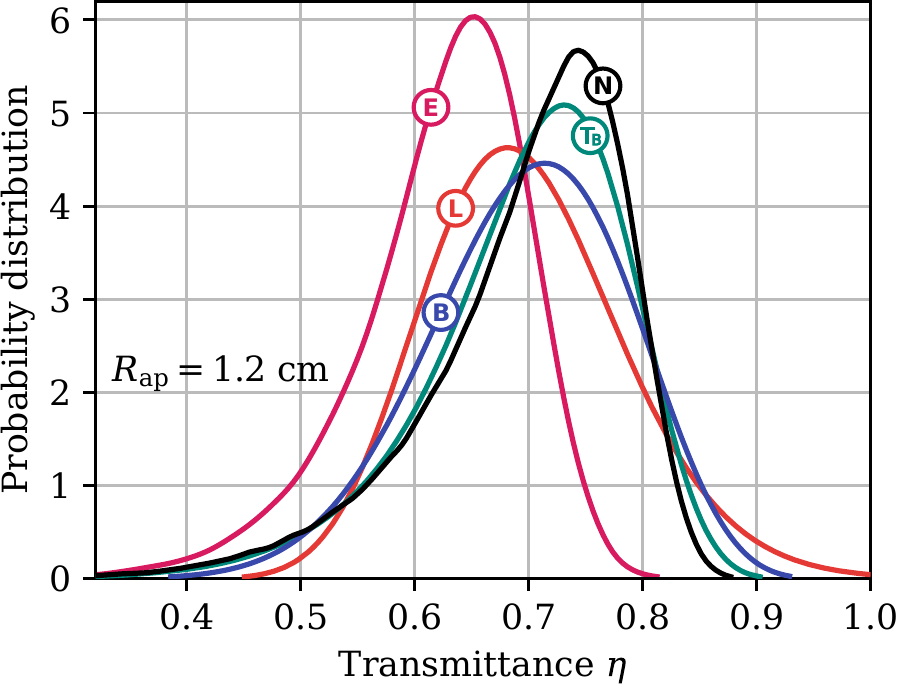}
			\caption{\label{Fig:weak_f_pdt} Numerically simulated (N) PDT, Beta (B) PDT, elliptic-beam (E) PDT, truncated log-normal (L) PDT, and the PDT based on the law of total probability with Beta distribution (T$_\textrm{B}$), shown for the channel with weak  impact of turbulence and  $F_0=z_\mathrm{ap}$.}
		\end{figure}

	The numerically simulated and analytical PDTs for the geometrically focused beam are shown in Fig.~\ref{Fig:weak_f_pdt}.
	It is interesting to note that the shape of the numerically simulated PDT is similar to the PDT for the elliptic-beam model, but for different values of the parameters.
	Consequently, the experimental or numerically simulated data can still be fitted by the elliptic-beam and beam-wandering models.
	However, the estimated values of the channel parameters obtained with such fitting are biased.
	In general, the modes of the elliptic-beam PDTs are less(greater) than the modes of the numerically simulated PDTs for aperture radii less(greater) than the value corresponding to the minimum in the KS statistic.

	It is worth noting that geometrical focusing significantly improves the channel characteristics under the considered conditions.
	This means that the effects of diffraction are still significant.
	In particular, the mean transmittance $\langle\eta\rangle$ is larger for $F_0=z_\mathrm{ap}$ than for $F_0=+\infty$.


\subsection{Channel with moderate  impact of turbulence}
\label{Sec:Moderate}

    The channel considered in this section (see the third column in Table~\ref{Tab:Regimes}) is characterized by the same beam parameters as the channel considered in Sec.~\ref{Sec:Weak}.
	However, its length is 1.6~km.
	Such a channel has been implemented in Erlangen, Germany (see Refs.~\cite{peuntinger14,usenko12,heim14}).
	In contrast to the previous case, here we consider stronger local turbulence, resulting in a moderate impact on the light beam.
	Since the propagation distance is approximately equal to the Rayleigh length, the corresponding beam is significantly divergent for $F_0=+\infty$.
	Therefore, the geometrically focused beam, $F_0=z_\mathrm{ap}$, is the optimal choice for given $W_0$ and $z_\mathrm{ap}$.

\begin{figure}[ht!]
                \includegraphics[width=1\linewidth]{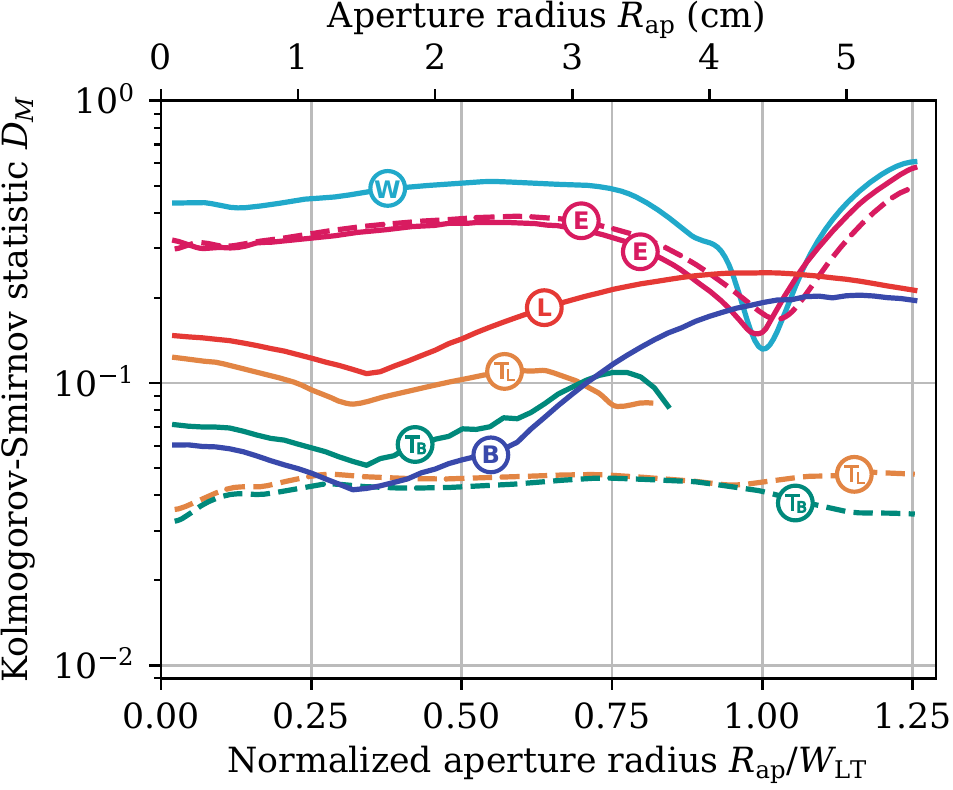}
			   	\caption{\label{Fig:erlangen_f_ks}  The KS statistics, quantifying differences between the numerically simulated and analytical or semianalytical PDTs, shown for the beam-wandering model (W), the elliptic-beam model (E), the truncated log-normal model (L), the model based on the law of total probability with the truncated log-normal (T$_\mathrm{L}$) and Beta (T$_\textrm{B}$) distribution, and the Beta-distribution model (B) for the channel with moderate impact of turbulence and $F_0=z_\mathrm{ap}$.
			   	Solid and dashed lines correspond to analytical and semi-analytical models, respectively.}
		   \end{figure}

    The corresponding KS statistics are shown in Fig.~\ref{Fig:erlangen_f_ks}.
	The best agreement with the numerically simulated data is achieved by the models based on the law of total probability.
	This fact becomes clear from the KS statistic for the corresponding semianalytical model.
	However, these KS statistics, similar to the previously considered case, suffer from a significant statistical error.
	The Beta-distribution model shows the best agreement among the analytical models for almost the entire range of aperture radius values.
	Exceptions are the beam-wandering and elliptic-beam models, which show the best result in a range with aperture radius close to the long-term beam-spot radius $W_\mathrm{LT}$.
	The latter models may still be capable of fitting numerically simulated data for a wider range of aperture radii.
	However, this may lead to a biased estimation of the turbulence parameters.

		    \begin{figure}[ht!]
		    	\includegraphics[width=1\linewidth]{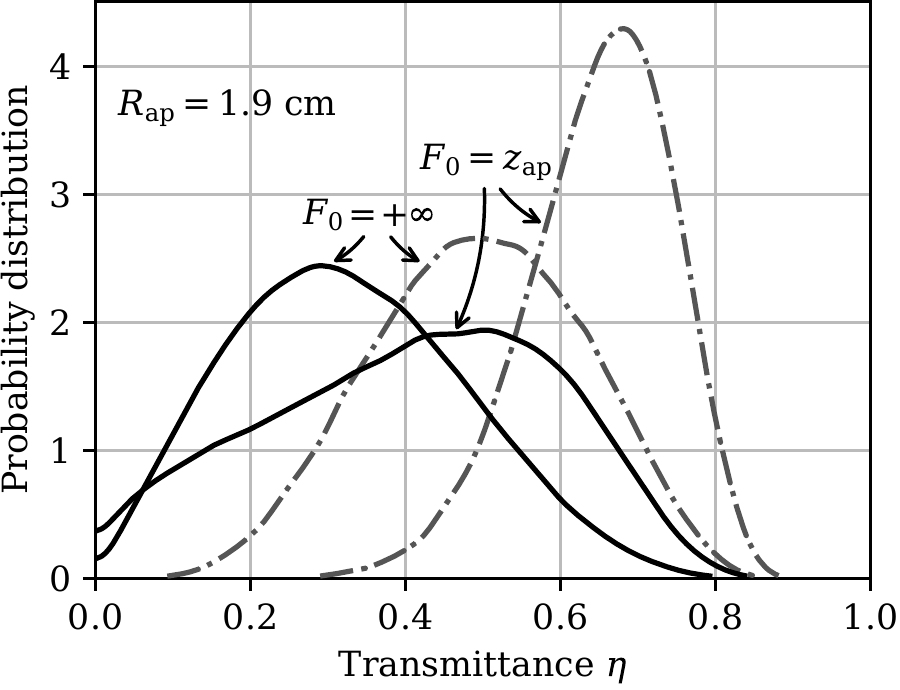}
		    	\caption{\label{Fig:erlangen_pdt} Numerically simulated PDTs for the channel with the moderate  impact of turbulence.
		    		The cases $F_0=z_\mathrm{ap}$ and $F_0=+\infty$ are given for the untracked (solid lines) and perfectly tracked (dot-dashed lines) beams.
		    	}
		    \end{figure}

    Examples of the numerically simulated PDTs are shown in Fig.~\ref{Fig:erlangen_pdt} for $F_0=z_\mathrm{ap}$ and $F_0=+\infty$.
	For the considered aperture radius $R_\mathrm{ap}=1.9~\mathrm{cm}$, the PDT shapes do not resemble the beam-wandering and/or elliptic-beam PDTs.
	We also consider a scenario with the beam-tracking procedure (see, e.g., Ref.~\cite{Kaushal_book}).
	The idea is to reduce the signal noise caused by beam wandering.
	A typical setup consists of a beacon laser at the receiver side and a position-sensitive sensor at the transmitter side.
	The latter is used for fast control of the source to align the beam centroid with the aperture center.
	The transformation of the PDT due to the partial and the complete mitigation of the beam wandering has been considered in Ref.~\cite{vasylyev18}.
	The PDTs for the perfectly tracked beams are shown in Fig.~\ref{Fig:erlangen_pdt}.
	These distributions describe only the beam-spot distortion.
	One can see that in this case, contrary to the assumptions in the model based on the law of total probability, these PDTs do not agree with the truncated log-normal distributions.
	It is also clear that the beam tracking procedure leads to an increase in the PDT modes.
	This implies that the channel characteristics are significantly improved with beam tracking.


\subsection{Channel with strong  impact of turbulence}
\label{sec:strong_turbulence_channel}

	In this section we consider a 50~km channel with weak strength of local turbulence (see the fourth column in Table~\ref{Tab:Regimes}).
	The propagation distance plays a crucial role in such a scenario; thus the overall impact of turbulence becomes strong.
	The Rayleigh length in this case is $z_\mathrm{R}\approx 14$~km.
	In the absence of atmosphere, the beam-spot radius at the aperture plane is almost the same for $F_0=z_\mathrm{ap}$ and   $F_0=+\infty$.
	Therefore, it is reasonable to consider only the latter case.

		\begin{figure}[ht!]
			\includegraphics[width=1\linewidth]{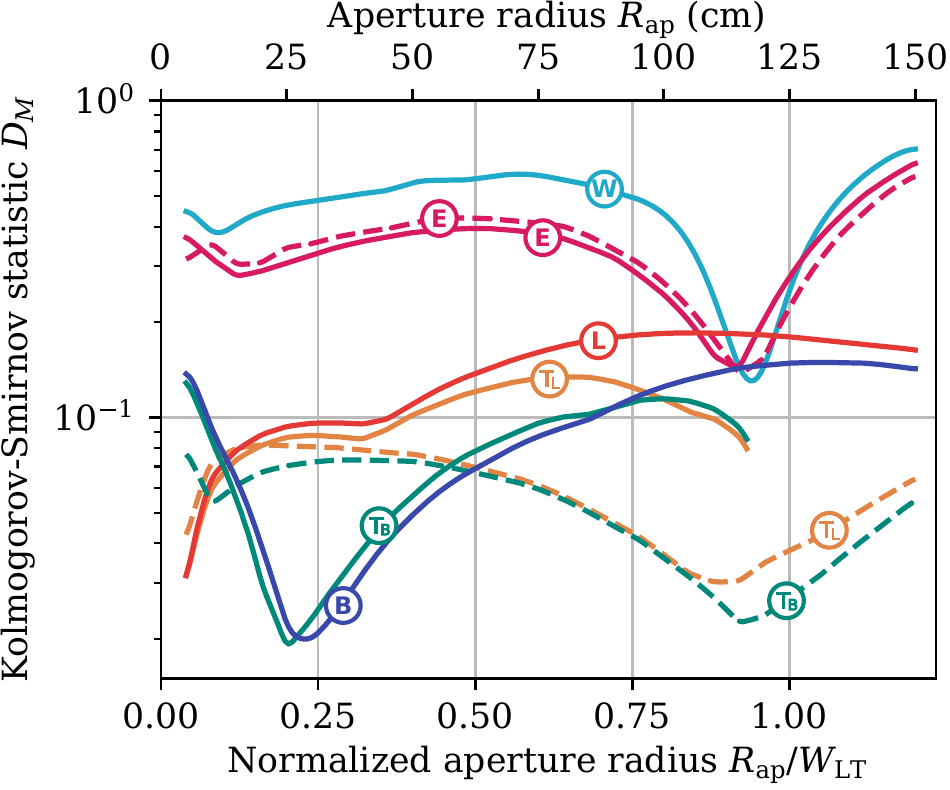}
			\caption{\label{Fig:viennian_f_ks} The KS statistics for the beam-wandering model (W), the elliptic-beam model (E), the truncated log-normal model (L), the model based on the law of total probability with the truncated log-normal (T$_\mathrm{L}$) and Beta (T$_\textrm{B}$) distributions, and the Beta-distribution model (B) for the channel with strong  impact of turbulence and  $F_0=z_\mathrm{ap}$.
			Solid and dashed lines correspond to analytical and semi-analytical models, respectively.}
		\end{figure}

		\begin{figure}[h!!!]
			\includegraphics[width=0.99\linewidth]{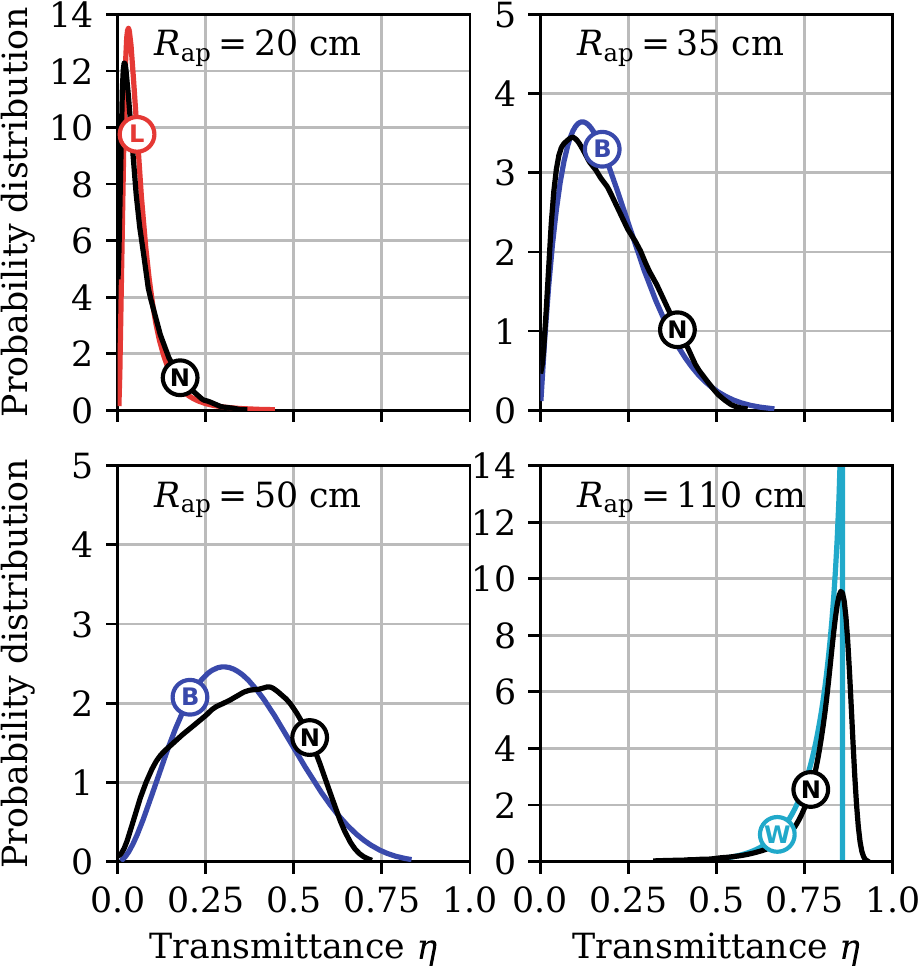}
			\caption{\label{Fig:viennian_f_pdt} Numerically simulated (N), truncated log-normal (L), Beta (B), and beam-wandering (W) PDTs for the channel with strong  impact of turbulence and $F_0=z_\mathrm{ap}$, shown for different values of the aperture radius.}
		\end{figure}

	The KS statistics, quantifying the differences between the numerically simulated data and the analytical or semianalytical models for the considered channel, are shown in Fig.~\ref{Fig:viennian_f_ks}.
	The general conclusions are similar to the case of moderate  impact of turbulence: The beam-wandering and elliptic-beam models show good agreement with the numerically simulated data for the aperture radius close to the long-term beam-spot radius $W_\mathrm{LT}$; outside this range, the best result is usually given by the Beta-distribution model.
	However, there is an important exception: For small apertures, the best result is given by the truncated log-normal model and the model based on the law of total probability with the truncated log-normal distribution.
	The KS statistic for the semi-analytical models based on the law of total probability demonstrates the best result for large apertures.

	The numerical PDT and the Beta PDT for different values of the aperture radius are shown in Fig.~\ref{Fig:viennian_f_pdt}.
	Obviously, the PDT skewness changes with increasing aperture; this property is inherent in all considered channels.
	For small apertures, the PDT mode is left centered such that its shape is close to the truncated log-normal distribution or the model based on the law of total probability.
	This can be explained by a leading role of the beam-spot distortion.
	For larger apertures, the finite size of the beam becomes important.
	As a consequence, the PDT resembles the distributions for the beam-wandering and/or elliptic-beam models, which are right centered.


\section{Statistical characteristics of beam parameters}
\label{Sec:Stat}

   The analytical and semianalytical models considered in the previous sections are based on assumptions about the statistical properties of the beam parameters.
   First, the beam-wandering model, the elliptic-beam model, and the model based on the law of total probability all assume that the beam-centroid position follows a two-dimensional Gaussian distribution.
   Then the model based on the law of total probability includes the assumption that the beam-centroid position is statistically independent of the beam shape.
   Finally, the elliptic-beam model supposes a particular shape of the probability distribution for the semiaxes of the ellipse, forming the beam shape in the Gaussian approximation.
   In this section we will verify these assumptions through numerical simulations.


   \subsection{Gaussianity of the beam-centroid position}

   Here we verify that the beam-centroid position follows a two-dimensional Gaussian distribution.
   To accomplish this, we calculate the beam centroid position using Eq.~(\ref{Eq:BeamCentroid}) for each sampling event and analyze the corresponding statistical data.
   Note that $x_0$ and $y_0$ are statistically independent.
   Therefore, it is sufficient to check whether a single coordinate follows a Gaussian distribution.

	   \begin{table}[h!]
	   	\caption{\label{Tab:BW} Skewness and excess kurtosis for the beam-centroid coordinate $x_0$ in the case of channels with weak, moderate, and strong  impact of turbulence, as given in Table~\ref{Tab:Regimes}.}
	   	\renewcommand{\arraystretch}{2.5}
	   	\begin{tabular}{lccc}
	   		\hline
	   		\hline
	   		\parbox{6em}{\raggedright Impact of\\  turbulence} & \parbox{5.5em}{Wave-front\\ radius $F_0$} & Skewness & \parbox{5.5em}{Excess\\ kurtosis}\\[1ex]
	   		\hline
	   		\multirow{2}*{weak} &$z_\mathrm{ap}$& $-0.0043$ & $-0.0038$ \\
	   		&$+\infty$& $0.0058$ & $0.014$ \\
	   		\multirow{2}*{moderate} &$z_\mathrm{ap}$&$-0.003$   &$-0.0279$  \\
	   		&$+\infty$& $0.0172$  & $-0.0046$ \\
	   		strong	&$+\infty$&  $0.0008$ &  $-0.1064$\\ \hline
	   	\end{tabular}
	   \end{table}

	   \begin{figure}[h!!!]
	   		\includegraphics[width=0.85\linewidth]{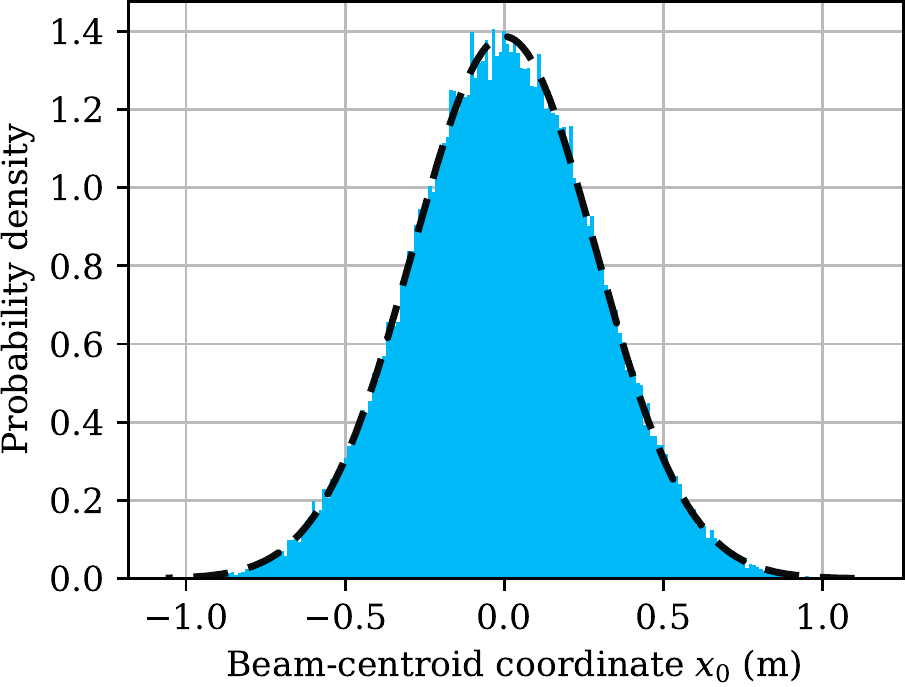}
		   	\caption{\label{Fig:bw_is_gauss} Histogram obtained from the numerically simulated data, demonstrating the distribution of the beam-centroid coordinate $x_0$ in the case of a channel with strong  impact of turbulence (cf. Table~\ref{Tab:Regimes}), compared with the corresponding Gaussian distribution (dashed line).}
	   \end{figure}

   To characterize non-Gaussianity, we will use skewness and excess kurtosis \cite{Joanes1998}.
   The corresponding values for different channels are given in Table~\ref{Tab:BW}.
   The largest discrepancy with the assumption of Gaussian distribution for the beam-centroid position is shown by the channel with strong  impact of turbulence.
   However, even in this case, the skewness and excess kurtosis are small.
   This implies that the numerically simulated data demonstrate good agreement with Gaussian distributions in all cases.
   In Fig.~\ref{Fig:bw_is_gauss} we compare numerically simulated data with the corresponding Gaussian distribution for the case where we find maximal absolute values of the skewness and excess kurtosis.
   As can be seen, the deviation from the Gaussian distribution is negligible even for such a scenario.


	\subsection{Statistical contribution of beam wandering}

	In this section we analyze the contribution of beam wandering into the PDT shape.
	This can be characterized by correlations between the beam-deflection distance $r_0=|\mathbf{r}_0|$ and the transmittance $\eta$.
	Non-zero values of the corresponding Pearson correlation coefficient
		\begin{align}\label{Eq:Corr_r0-eta}
			S(r_0,\eta)=\frac{\left\langle\Delta r_0 \Delta\eta\right\rangle}{\sqrt{\left\langle\Delta r_0^2\right\rangle\left\langle \Delta\eta^2\right\rangle}}
		\end{align}
	indicate a contribution of beam wandering into the PDT.

		\begin{figure}[ht!]
		\includegraphics[width=1\linewidth]{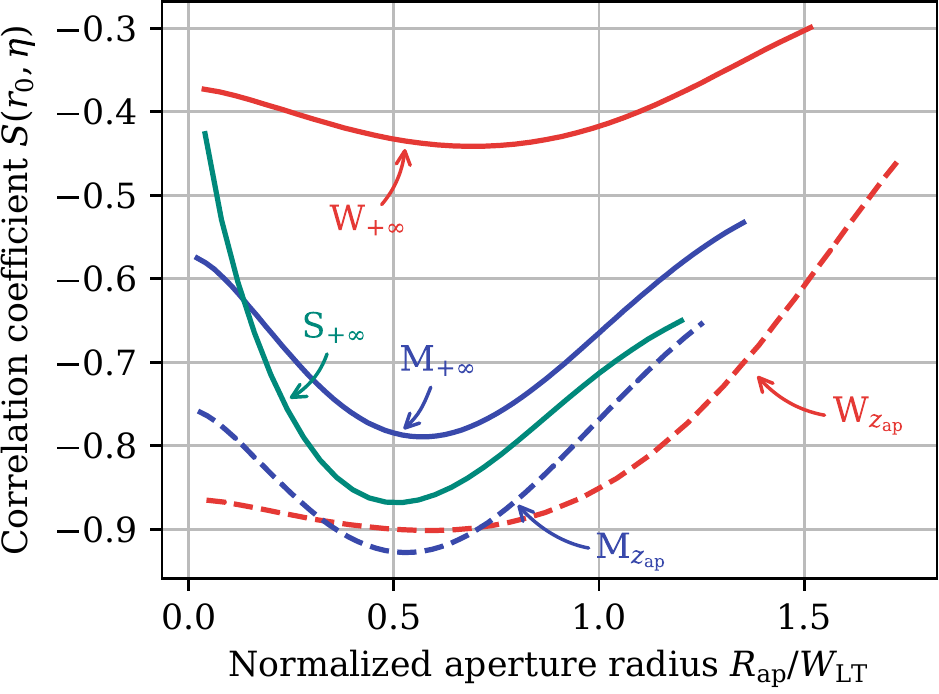}
		\caption{\label{Fig:Pearson_eta-r0-1} Pearson correlation coefficient (\ref{Eq:Corr_r0-eta}) as a function of aperture radius in units of the long beam-spot width corresponding to each scenario.
		Here $W$, $M$, and $S$ correspond to the channels with weak, moderate, and strong  impact of turbulence, respectively (see Table~\ref{Tab:Regimes}).
		The indices $+\infty$ and $z_{\mathrm{ap}}$ indicate the cases with $F_0=+\infty$ and $F_0=z_{\mathrm{ap}}$, respectively.}
		\end{figure}

	We analyze each sampling event and find the corresponding values of $\eta$ and $r_0$ for it by using Eqs.~(\ref{Eq:Efficiency}) and (\ref{Eq:BeamCentroid}), respectively.
	With this set of sampled data, we can estimate the Pearson correlation coefficient.
	The results are shown in Fig.~\ref{Fig:Pearson_eta-r0-1} as a function of the aperture radius in units of the corresponding long beam-spot width $R_{\mathrm{ap}}/W_{\mathrm{LT}}$.
	Obviously, the contribution of beam wandering in the PDT strictly depends on the aperture radius.
	The maximum anticorrelations appear for $R_{\mathrm{ap}}\approx 0.5 W_{\mathrm{LT}}$ for all channels.
	Also channels with weak  impact of turbulence and with $F_0=+\infty$ show relatively small anticorrelations compared to other cases.
	Strong anticorrelations shown by the Pearson coefficient (\ref{Eq:Corr_r0-eta}) indicate the need to explicitly account for the contribution of beam wandering such as it is done in the elliptic-beam model and in the model based on the law of total probability.


	\subsection{Correlations between beam shape and beam deflection}
	\label{Sec:Corr-Shape-Discr}

	The model based on the law of total probability \cite{vasylyev18} includes the assumption that the beam shape is independent of the beam deflection.
	This implies that $I^{(c)}(\mathbf{r})$ in Eq.~(\ref{Eq:Ic}) is statistically independent on $\mathbf{r}_0$.
	This assumption will be verified in this section.

	A straightforward way to do this is to study the correlations between the beam deflection $r_0$ and the transmittance of the perfectly tracked beam obtained via integration similar to Eq.~(\ref{Eq:Efficiency}) but for $I^{(c)}(\mathbf{r})$.
	For each sampling event, we find the corresponding beam-centroid position $\mathbf{r}_0$ using Eq.~(\ref{Eq:BeamCentroid}), place the aperture center there, and calculate the corresponding transmittance $\eta$ using Eq.~(\ref{Eq:Efficiency}).
	The result is the sample set of two random variables, $r_0=|\mathbf{r}_0|$ and $\eta$, which we use to estimate the Pearson correlation coefficient similar to Eq.~(\ref{Eq:Corr_r0-eta}).
	If it differs from zero, then the assumption of statistical independence between the beam shape and the beam deflection fails.

		\begin{figure}[ht!]
			\includegraphics[width=1\linewidth]{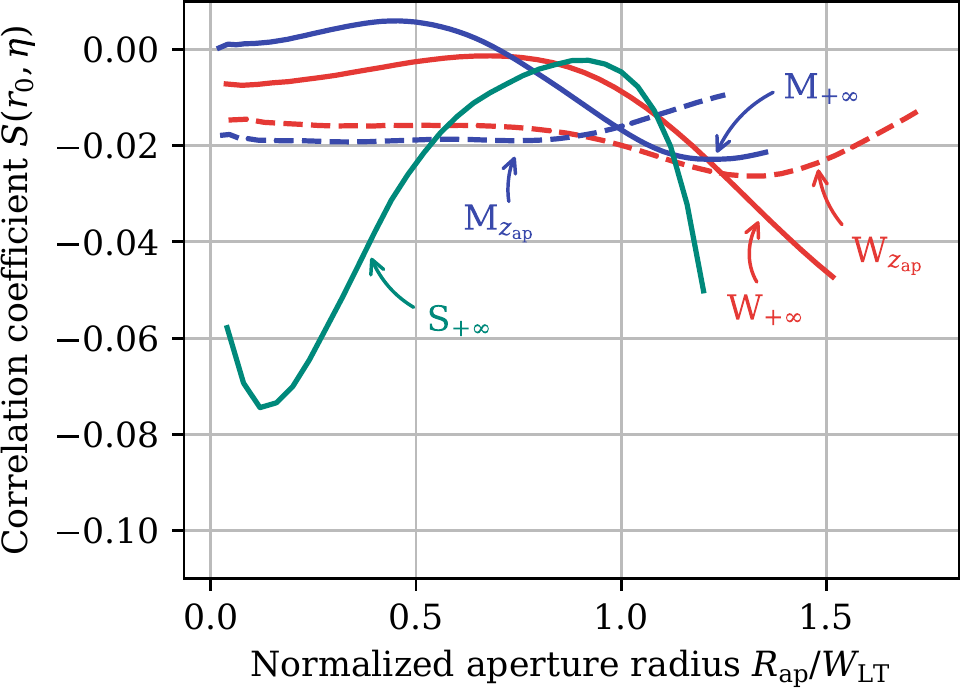}
			\caption{\label{Fig:Pearson_eta-r0-2} Pearson correlation coefficient, defined similarly to Eq.~(\ref{Eq:Corr_r0-eta}) but with the transmittance of the perfectly tracked beam, as a function of aperture radius in units of the long beam-spot width corresponding to each scenario.
			Lines are marked as in Fig.~\ref{Fig:Pearson_eta-r0-1}.}
		\end{figure}

	Values of this coefficient as a function of aperture radius are shown in Fig.~\ref{Fig:Pearson_eta-r0-2}.
	In most cases the transmittance of the perfectly tracked beam and the beam-deflection distance are only weakly anticorrelated at least in the Gaussian approximation.
	Relatively strong anticorrelations occur only in the case of channels with strong  impact of turbulence.

	    \begin{figure}[ht!]
			\includegraphics[width=0.6\linewidth]{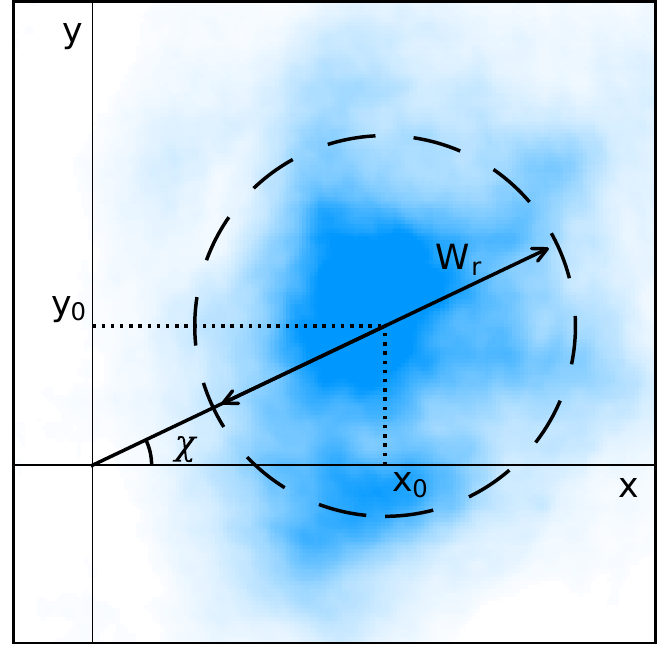}
			\caption{\label{Fig:img_corr_r0_wr} Tangential beam-spot width $W_r$.
			The  beam centroid is deflected to the point $\mathbf{r}_0=(x_0,y_0)$.
			The angle $\chi$ defines the direction of the vector $\mathbf{r}_0$ to the $x$ axis.}
		\end{figure}

		\begin{table}[h!]
			\caption{\label{Tab:Corr} Pearson correlation coefficient between the beam-deflection distance $r_0$ and the radial beam width $W_r$ [cf. Eq.~(\ref{Eq:Corr_rW})], for the channels with weak, moderate, and strong  impact of turbulence, as given in Table~\ref{Tab:Regimes}.}
			\renewcommand{\arraystretch}{2.5}
			\begin{tabular}{lcc}
				\hline
				\hline
				 Impact of turbulence & $F_0=+\infty$ & $F_0=z_\mathrm{ap}$\\[1ex]
				\hline
				\parbox[c]{3.35cm}{\raggedright weak} & $0.016$ & $0.039$ \\
				\parbox[c]{3.35cm}{\raggedright moderate} & $0.080$ & $0.150$ \\
				\parbox[c]{3.35cm}{\raggedright strong} & $0.317$ &  \\ \hline
			\end{tabular}
		\end{table}

	We also consider another way to verify statistical independence between the beam-deflection distance and the beam shape, which does not depend on the aperture.
	We characterize the beam shape using  the tangential beam-spot width $W_r$ as depicted in Fig.~\ref{Fig:img_corr_r0_wr}.
	For each sampling event, we find the beam-centroid position $\mathbf{r}_0=(x_0,y_0)$ and the corresponding angle $\chi$ that defines the direction of the vector $\mathbf{r}_0$ to the $x$ axis such that $\tan\chi=x_0/y_0$. In the coordinate frame $(x_r,y_r)$, oriented to the beam centroid,
		\begin{align}
			&x_r=x\cos\chi+y\sin\chi,\\
			&y_r=-x\sin\chi+y\cos\chi,
		\end{align}
	we calculate $S_{x_rx_r}$ similarly to Eq.~(\ref{Eq:MatrixS_Definition}) and the corresponding beam width as $W_r = \sqrt{S_{x_rx_r}}$.
	From the set of sampling events, we obtain the Pearson correlation coefficient between $W_r$ and $r_0=(x_0^2+y_0^2)^{1/2}$,
        \begin{align}\label{Eq:Corr_rW}
			S(r_0,W_r)=\frac{\langle \Delta r_0\Delta W_r \rangle}{\sqrt{\langle\Delta r_0 ^2\rangle\langle \Delta W_r^2\rangle }},
		\end{align}
	characterizing statistical correlations between the beam centroid and the beam shape.
	The results for different channels are shown in Table~\ref{Tab:Corr}.
	They demonstrate that these statistical correlations can be significant for channels with moderate and strong impact of turbulence.
	However, as seen above, for a wide range of aperture radii this correlation does not significantly affect the transmittance values.


    \subsection{Statistical distribution of semi-axes, forming elliptic beams}

   Let us consider the semiaxes of the ellipse, forming the beam shape in the Gaussian approximation.
   These quantities can be defined as eigenvalues of the matrix $\mathbf{S}$ [cf. Eq.~(\ref{Eq:MatrixS_Definition})],
	    \begin{align}
		    W_{1,2}^2 =\frac{1}{2} \left[S_{xx} + S_{yy} \pm \sqrt{(S_{xx} - S_{yy})^2 + 4S_{xy}^2}\right],
		    \nonumber
	    \end{align}
	such that $W_1$ is assigned the eigenvalue with the sign "$+$" or "$-$" if $S_{xy}\geq0$ or $S_{xy}\leq0$, respectively.
    As it is assumed in the elliptic-beam model \cite{vasylyev16}, the quantities
	    \begin{align}\label{Eq:Theta}
	    	\Theta_{1,2}=\ln W_{1,2}^2/W_0^2
	    \end{align}
    are distributed according to a bivariate Gaussian distribution.
    This assumption is verified in this section.

\begin{figure}[ht!]
     		\includegraphics[width=0.7\linewidth]{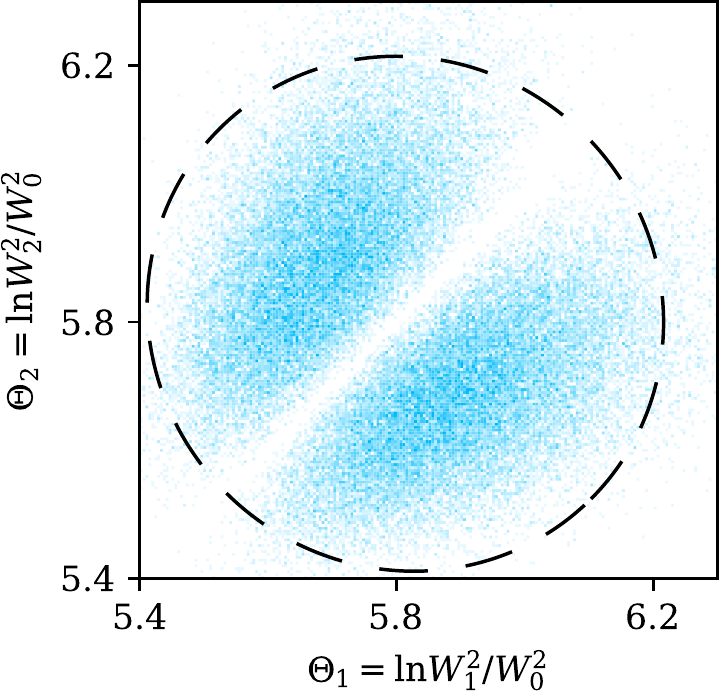}
     		\caption{\label{Fig:theta1_theta2} Coarse-grained scatter plot for $\Theta_1$ and $\Theta_2$ in the elliptic-beam model for the channel with  strong  impact of turbulence.
     		The dashed line corresponds to the covariance ellipse, described by Eq.~(\ref{Eq:CovEllipce}).}
     	\end{figure}

    For each sampling event we numerically calculate the matrix $\mathbf{S}$ according to Eq.~(\ref{Eq:MatrixS_Definition}) and the corresponding values of $\Theta_1$ and $\Theta_2$.
    The corresponding coarse-grained scatter plot for the channel with strong  impact of turbulence and $F_0=z_{\mathrm{ap}}$ is shown in Fig.~\ref{Fig:theta1_theta2}.
    It is compared with the covariance ellipse, given by the equation
    	\begin{align}\label{Eq:CovEllipce}
    		\sum\limits_{i,j=1}^2\big(\Theta_i-\langle\Theta_i\rangle\big)\Sigma_{ij}^{-1}\big(\Theta_j-\langle\Theta_j\rangle\big)=4,
    	\end{align}
	where $\Sigma_{ij}=\left\langle \Delta\Theta_i\Delta\Theta_j \right\rangle$ is the covariance matrix for the variables  $\Theta_1$ and $\Theta_2$.
   	Obviously, the corresponding distribution differs significantly from the Gaussian distribution, especially due to the pronounced minimum for $\Theta_1=\Theta_2$.
    Another important observation is that the Pearson correlation coefficient for these quantities is small; the variance ellipse almost resembles a circle.
    Consequently, $\Theta_1$ and $\Theta_2$ can be considered as uncorrelated in the second order.

	     \begin{table}[h!]
	    	\caption{\label{Tab:Theta} Skewness and excess kurtosis for $\Theta_\mathrm{(c)}$ and $\Theta_\mathrm{(s)}$ in the case of channels with weak, moderate, and strong  impact of turbulence, as given in Table~\ref{Tab:Regimes}.}
	    	\renewcommand{\arraystretch}{2.5}
	    	\begin{tabular}{lccc}
	    		\hline
	    		\hline
	    		\parbox{4.3em} {\raggedright Impact of\\  turbulence} & \parbox{5.5em}{Wave-front\\ radius $F_0$} & \parbox{7.3em}{Skewness\\ for $\Theta_{(\mathrm{c})}$ and $\Theta_{(\mathrm{s})}$} & \parbox{7.5em}{Excess kurtosis\\for $\Theta_{(\mathrm{c})}$ and $\Theta_{(\mathrm{s})}$}\\[1ex]
	    		\hline
	    		\multirow{2}*{weak} &$z_\mathrm{ap}$& \parbox{6.5em}{$2.5\times10^{-1}$\\$-6.1\times10^{-4}$} & \parbox{6.5em}{$2.3\times10^{-1}$\\$-8.5\times10^{-1}$} \\
	    		&$+\infty$& \parbox{6.5em}{$-9.9\times10^{-2}$\\$-1.0\times10^{-2}$} & \parbox{6.5em}{$-6.4\times10^{-3}$\\$-1.0$} \\
	    		\multirow{2}*{moderate} &$z_\mathrm{ap}$&\parbox{6.5em}{$2.5\times10^{-1}$\\$9.8\times10^{-3}$}   &\parbox{6.5em}{$1.2\times10^{-1}$\\$-8.1\times10^{-1}$}  \\
	    		&$+\infty$& \parbox{6.5em}{$1.5\times10^{-1}$\\$-3.7\times10^{-3}$}  & \parbox{6.5em}{$-1.1\times10^{-1}$\\$-9.9\times10^{-1}$} \\
	    		strong&$+\infty$&  \parbox{6.5em}{$3.2\times10^{-1}$\\$-9\times10^{-3}$} &  \parbox{6.5em}{$2.5\times10^{-1}$\\$-7.7\times10^{-1}$}\\ \hline
	    	\end{tabular}
	    \end{table}

	In order to characterize non-Gaussianity, we first rotate the coordinate system $(\Theta_1,\Theta_2)$ to another one, $(\Theta_\mathrm{(c)},\Theta_\mathrm{(s)})$, as
		\begin{align}
			&\Theta_\mathrm{(c)}=\frac{1}{\sqrt{2}}\left(\Theta_1+\Theta_2\right),\\
			&\Theta_\mathrm{(s)}=\frac{1}{\sqrt{2}}\left(\Theta_1-\Theta_2\right).
		\end{align}
	Since the probability distribution function is symmetrical with respect to the axis $\Theta_\mathrm{(c)}$, the quantities $\Theta_\mathrm{(c)}$ and $\Theta_\mathrm{(s)}$ should be uncorrelated.
	We then calculate the skewness and excess kurtosis for these quantities in order to conclude about non-Gaussianity of the probability distribution for $\Theta_1$ and $\Theta_2$.
	The results are summarized in Table~\ref{Tab:Theta}.
	Significant deviations from the Gaussian distribution indicate the further need to include this feature in the elliptic-beam model.


\section{Application: Transmission of quadrature-squeezed light}
\label{Sec:Appl}

	Quadrature squeezing is a fundamental example of nonclassical phenomena exhibited by quantum light when measuring a field quadrature $\hat{x}=2^{-1/2}(\hat{a}+\hat{a}^\dag)$, where $\hat{a}$ is the field annihilation operator \cite{walls83,Slusher1985,Wu1986}.
	The quadrature variances can be expressed as
		\begin{align}
			\left\langle\Delta \hat{x}^2 \right\rangle=\frac{1}{2}+\left\langle:\Delta \hat{x}^2: \right\rangle,
		\end{align}
	where the term $\frac{1}{2}$ corresponds to the quadrature variance of the vacuum state and $\left\langle:\Delta \hat{x}^2: \right\rangle$ represents the normal-ordered quadrature variance.
	A negative value for $\left\langle:\Delta \hat{x}^2: \right\rangle$ indicates that the quadrature variance is smaller than its vacuum-state value.

	A scheme of quadrature measurement with homodyne detection for the light passing through the turbulent atmosphere has been proposed and implemented in Refs.~\cite{elser09,heim10}.
	In this case, the local oscillator is transmitted in an orthogonally polarized but the same spatial mode as the signal.
	Since the depolarization effect of the atmosphere is negligible (see, e.g., Ref.~\cite{Tatarskii}), the transmittance of the local oscillator and the signal are strongly correlated.
	This correlation provides an opportunity to control the value of channel transmittance and postselect the events with high transmittance.
	Such a scheme has been implemented in Ref.~\cite{peuntinger14} and theoretically analyzed in Refs.~\cite{semenov12,vasylyev16}.

	As it has been shown in Ref.~\cite{semenov12}, the normal-ordered quadrature variance at the receiver site, $\left\langle:\Delta \hat{x}^2: \right\rangle_\mathrm{out}$, is related to the  normal-ordered quadrature variance at the transmitter site, $\left\langle:\Delta \hat{x}^2: \right\rangle_\mathrm{in}$, as
		\begin{align}\label{Eq:IOR_Sq}
			\left\langle:\Delta \hat{x}^2: \right\rangle_\mathrm{out}=\left\langle\eta\right\rangle\left\langle:\Delta \hat{x}^2: \right\rangle_\mathrm{in}
			+\left\langle\Delta T^2\right\rangle\left\langle\hat{x}\right\rangle_\mathrm{in}^2,
		\end{align}
	where $T=\sqrt{\eta}$ is the transmission coefficient.
	It is important to note that the constant losses on absorption and scattering in the atmosphere as well as the losses of the optical system should be accounted for in the PDT or in the numerically simulated data.
	For the squeezed vacuum state one has $\left\langle\hat{x}\right\rangle_\mathrm{in}=0$ and only the first term in Eq.~(\ref{Eq:IOR_Sq}) plays a role.
	The mean transmittance for the postselected signal is given by
		\begin{align}
			\left\langle\eta\right\rangle=\frac{1}{\overline{\mathcal{F}}(\eta_\mathrm{min})}
			\int\limits_{\eta_\mathrm{min}}^{1}d\eta\,\eta\,\mathcal{P}\left(\eta\right),
		\end{align}
	where $\eta_\mathrm{min}$ is the minimal value of the postselected efficiency (the postselection threshold) and $\overline{\mathcal{F}}(\eta_\mathrm{min})=\int_{\eta_\mathrm{min}}^{1}d\eta\,\eta\,\mathcal{P}\left(\eta\right)$ is the exceedance.
	The same quantity can be evaluated from the numerically simulated data as
		\begin{align}
			\left\langle\eta\right\rangle=\frac{1}{N_\mathrm{min}}\sum\limits_{\{\eta_i\geq \eta_{\mathrm{min}}\}} \eta_i,
		\end{align}
	where $\eta_i$ are sampled values of the transmittance values and $N_\mathrm{min}$ is the number of events with $\eta_i\geq \eta_{\mathrm{min}}$.

\begin{figure}[ht!]
				\includegraphics	[width=1\linewidth]{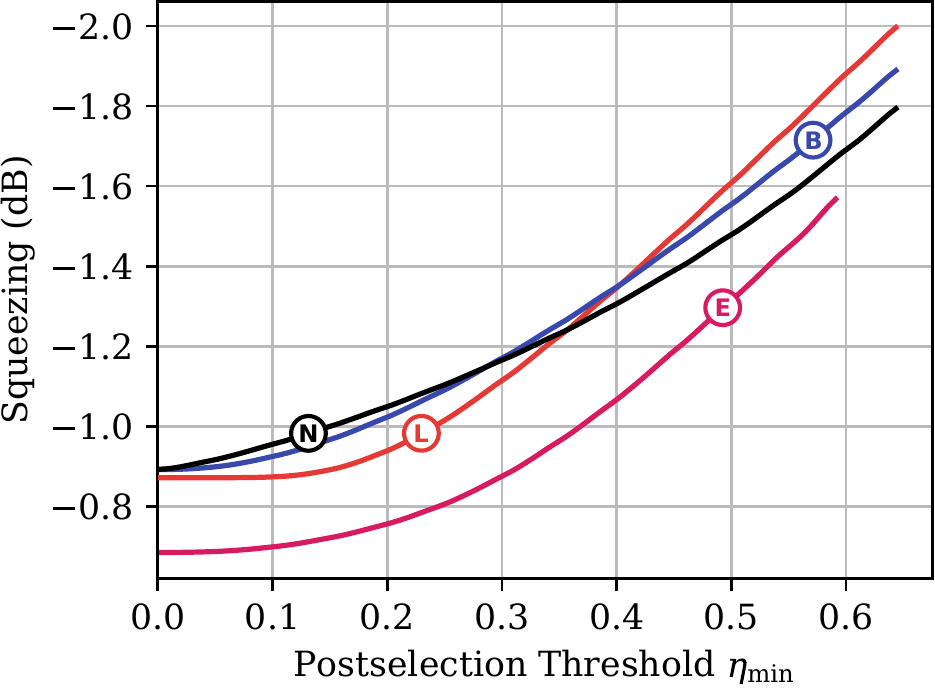}
				\caption{\label{Fig:SqEta} Squeezing parameter at the receiver as a function of the postselection threshold $\eta_{\mathrm{min}}$, evaluated with the numerically simulated data ($\mathrm{N}$), the elliptic-beam approximation ($\mathrm{E}$), the truncated log-normal distribution ($\mathrm{L}$), and the Beta-distribution model ($B$), shown for the channel with moderate  impact of turbulence (cf. Table~\ref{Tab:Regimes}) and $F_0=z_\mathrm{ap}$.
				The initial squeezing at the transmitter site is $-3$~dB.
				The constant losses are $0.38$~dB.
				The receiver aperture radius is $R_\mathrm{ap}=0.43W_\mathrm{LT}=0.019~\mathrm{cm}$.
			}
		\end{figure}

	We consider quadrature squeezing for the channels listed in Table~\ref{Tab:Regimes}.
	A typical example of the dependence of the squeezing parameter on the postselection threshold $\eta_\mathrm{min}$ is shown in Fig.~\ref{Fig:SqEta} for the channel with moderate  impact of turbulence and $F_0=z_\mathrm{ap}$.
	A property common to all the examples considered is that the elliptic-beam and beam-wandering models show a significant discrepancy with the numerically simulated data in the case of a small postselection threshold $\eta_\mathrm{min}$.
	This is due to the fact that these models are not bound to the value of the first moment $\langle\eta\rangle$.
	Another approach \cite{vasylyev16,vasylyev17} assumes an estimation of the channel parameters by the method of moments based on the beam-wandering and elliptic-beam PDTs.
	This leads to good agreement with experimental (numerically-simulated) data.
	However, this method leads to biased values of the estimated parameters of channels.

		\begin{figure}[ht!]
			\includegraphics[width=1\linewidth]{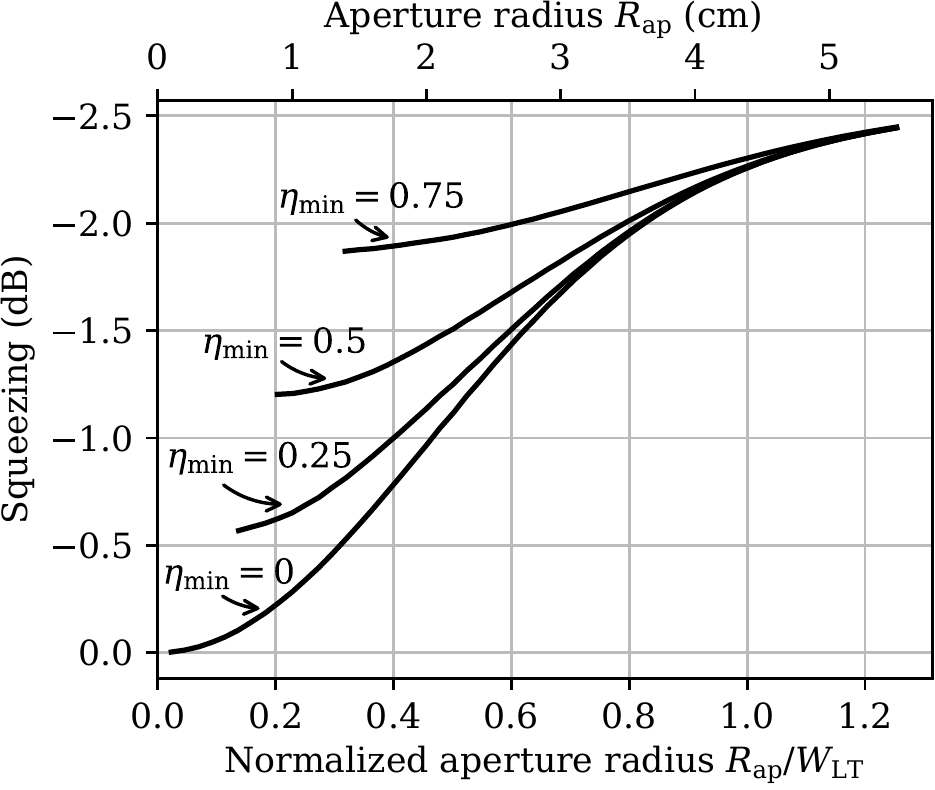}
			\caption{\label{Fig:SqApp} Squeezing parameter at the receiver as a function of the aperture radius $R_\mathrm{ap}$ shown for different values of the postselection thresholds $\eta_\mathrm{min}$. The channel parameters, the squeezing at the transmitter, and the constant losses are the same as in Fig.~\ref{Fig:SqEta}.}
		\end{figure}

	Another interesting observation from Fig.~\ref{Fig:SqEta} is a slight discrepancy between the squeezing parameters obtained from the truncated log-normal distribution and the numerically-simulated data at low postselection thresholds.
	This discrepancy arises due to the influence of the tail cutoff on the first two moments of the transmittance, leading to an error in Eqs.~(\ref{Eq:Mu}) and (\ref{Eq:Sigma}).
	The semianalytical models based on the law of total probability also exhibit a small discrepancy due to the correlations between beam shape and beam deflection discussed in Sec.~\ref{Sec:Corr-Shape-Discr}.
	However, analytical models based on the law of total probability with the approximation of small beam wandering do not exhibit this discrepancy since this approximation assumes a strong constraint on the first two moments of the transmittance.

	The Beta-distribution model shows the best agreement with the numerically simulated data.
	The corresponding PDT is strictly constrained to the first two moments of the transmittance, making this model work well for small values of the postselection threshold.
	However, this model also gives acceptable results for larger values of $\eta_\mathrm{min}$.

	The dependence of the squeezing parameters on the aperture radius is shown in Fig.~\ref{Fig:SqApp} for different values of the postselection threshold $\eta_\mathrm{min}$.
	The behavior of these parameters is generally predictable.
	For instance, a small value of the aperture radius may require a large value of the postselection threshold to achieve an acceptable value of the squeezing parameter.
	However, as the aperture radius approaches the long-term beam-spot radius, the postselection becomes less significant.


\section{Conclusions}
\label{Sec:Concl}

	We have presented numerical simulations of the PDT for horizontal  atmospheric quantum channels.
	Depending on the Rytov parameter values, we considered scenarios with weak, strong, and moderate  impact of turbulence.
	We performed numerical simulations using the sparse-spectrum model for the phase-screen method and compared our results with existing analytical models, including the truncated log-normal distribution, beam-wandering model, elliptic-beam approximation, and the model based on the law of total probability.
	Additionally, we compared our results with the empirical model presented here based on the Beta distribution and a corresponding modification of the model based on the law of total probability.
	We also tested nontrivial statistical assumptions on beam parameters at the receiver underlying analytical models of the PDT.

	An important conclusion drawn from this study is that the applicability of analytical models depends strongly on the aperture radius.
	At the same time, the dependence on the turbulence strength has a significantly smaller effect on this applicability.
    For instance, both the beam-wandering and elliptic-beam models demonstrate good agreement with numerically simulated data for aperture radii of the order of or slightly exceeding the long-term beam-spot radius.
	If the aperture radius is smaller or larger than this value, the PDT shape is similar to beam-wandering and elliptic-beam PDT, but with the left- and right-hand-side shifted distribution modes, respectively.
   This implies that the elliptic-beam model may fit numerically simulated data well within a wider range of aperture-radius values.
    However, such a fitting may result in biased estimations of turbulence parameters.

	The empirical Beta-distribution model presented here, as well as the corresponding model based on the law of total probability, has shown the best agreement with numerically simulated data over a wide range of aperture radii.
	The Beta-distribution model has a simple analytical form and its parameters depend solely on the first and second moments of the transmittance.
	It should be noted that in practical scenarios, determining the range of applicability of the beam-wandering and elliptic-beam models with precision may be difficult.
	Therefore, the use of the Beta-distribution model or the corresponding model based on the law of total probability may still be relevant, even if the aperture radius is expected to fall within this range.
	Considering this, the Beta-distribution model can be regarded as highly applicable for theoretical research on atmospheric quantum channels.

	We have applied our results to consider the transfer of quadrature squeezing through atmospheric channels.
	Proper experimental design allows for monitoring the actual transmittance value and postselecting events with high transmittance.
	The resulting squeezing depends on the mean transmittance for the postselected data.
	However, the beam-wandering and elliptic-beam models usually do not accurately approximate $\langle\eta\rangle$, resulting in worse agreement with numerically simulated data for small values of the postselection threshold than other models.
	Therefore, in scenarios where the accuracy of both $\langle\eta\rangle$ and $\langle\eta^2\rangle$ is crucial, the Beta-distribution model proves to be a more reliable solution.
	It demonstrates better agreement with the numerically simulated data, particularly for small minimum values of the postselected transmittance, than the beam-wandering and elliptic-beam models.

	This work was supported by the National Research Foundation of Ukraine through Project No. 2020.02/0111, Nonclassical and hybrid correlations of quantum systems under realistic conditions.
	The authors thank D. Vasylyev and M. Bohmann for insightful discussions.

\appendix

\section{Input-output relations}
\label{App:IOR}

    In this appendix we briefly remind reader of the derivation of the input-output relation~(\ref{Eq:POutDist}) and the expression for the transmittance $\eta$ given by Eq.~(\ref{Eq:Efficiency}).
    More details on the derivation of these expressions can be found in Appendixes~A, B, and C of the Supplemental Material in Ref.~\cite{vasylyev12}.
    As mentioned in Introduction, this consideration applies only to transmission protocols with fixed spatial structure of light modes at the transmitter.

	Let us consider a quasimonochromatic light mode described by the function
		\begin{align}
			V_{\mathrm{in}}\left(\mathbf{r},z;t\right)=
			\int_{\mathbb{R}}d k \Phi(k) u_{\mathrm{in}}\left(\mathbf{r},z;k\right)e^{ik(z-ct)}.
		\end{align}
	Here $\Phi(k)$ is a normalized spectrally narrow function, $u_{\mathrm{in}}\left(\mathbf{r},z;k\right)$ is a solution to the paraxial equation (\ref{Eq:Paraxial}) with a fixed $\delta n(\mathbf{r},z)$, and $c$ is the speed of light.
	Since $u_{\mathrm{in}}\left(\mathbf{r},z;k\right)$ varies much slower with $z$ than $e^{ikz}$, the modes with orthonormal time-spectrum functions can be considered as approximately orthonormal with respect to the $L_2$ scalar product in $\mathbb{R}^3$.

	The light at the transmitter side is prepared in a quantum state of the quasi-monochromatic mode $V_{\mathrm{in}}\left(\mathbf{r},z;t\right)$, which is described by the $P$ function
		\begin{align}\label{Eq:PIn}
			P_{\mathrm{in}}(\alpha)=\frac{1}{\pi^2}\int_{\mathbb{C}}d^2 \beta \Tr\left[\hat{\rho}\,e^{(\hat{a}^\dag_{\mathrm{in}}-\alpha^\ast)\beta}e^{-(\hat{a}_{\mathrm{in}}-\alpha)\beta^\ast}\right].
		\end{align}
	Here $\hat{\rho}$ is the density operator and the annihilation operator $\hat{a}_{\mathrm{in}}$ is expressed in terms of the positive-frequency part of the electromagnetic-field operator $\hat{A}_{\mathrm{in}}^{(+)}(\mathbf{r},z;t)$ as
		\begin{align}
			\hat{a}_{\mathrm{in}}=\int_{\mathbb{R}^3}d^2{\mathbf{r}}dz\hat{A}_{\mathrm{in}}^{(+)}(\mathbf{r},z;t) 						V_{\mathrm{in}}^{\ast}\left(\mathbf{r},z;t\right).
		\end{align}
	For brevity, we consider a scalar field, i.e., polarized light, and the dimensionless field operator.

	We also consider another (unnormalized) solution to Eq.~(\ref{Eq:Paraxial}),
		\begin{align}
			u_{\mathrm{out}}\left(\mathbf{r},z;k\right)=\int_{\mathbb{R}^2}d^2\mathbf{r}^\prime T(\mathbf{r},\mathbf{r}^\prime;z;z_{\mathrm{ap}})u_{\mathrm{in}}\left(\mathbf{r}^\prime,z;k\right).
		\end{align}
	The integral kernel $T(\mathbf{r},\mathbf{r}^\prime;z;z_{\mathrm{ap}})$ can be uniquely defined from the boundary condition at the aperture plane,
		\begin{align}\label{Eq:Uout}
			u_{\mathrm{out}}\left(\mathbf{r},z_{\mathrm{ap}};k\right)=I_{\mathcal{A}}(\mathbf{r}) 		u_{\mathrm{in}}\left(\mathbf{r},z_{\mathrm{ap}};k\right),
		\end{align}
	where $I_{\mathcal{A}}(\mathbf{r})$ is the indicator function equal to unity inside the aperture opening $\mathcal{A}$ and zero outside.
	This implies that after passing through the aperture, the normalized quasimonochromatic mode reads
		\begin{align}\label{Eq:Vout}
			 &V_{\mathrm{out}}\left(\mathbf{r},z;t\right)\nonumber\\
			&=\frac{1}{\sqrt{\eta}}
			\int_{\mathbb{R}}d k \Phi(k) u_{\mathrm{out}}\left(\mathbf{r},z;k\right)e^{ik(z-ct)},
		\end{align}
	where the transmittance $\eta$ is defined from the normalization condition of the function $V_{\mathrm{out}}\left(\mathbf{r},z;t\right)$ in $\mathbb{R}^3$.
	Similarly, the expression
		\begin{align}\label{Eq:IOR_Position}
			&\hat{A}_{\mathrm{out}}^{(+)}(\mathbf{r},z;t)\\
			&=\int_{\mathbb{R}^2}d^2\mathbf{r}^\prime T(\mathbf{r},\mathbf{r}^\prime;z;z_{\mathrm{ap}})\hat{A}_{\mathrm{in}}^{(+)}(\mathbf{r}^\prime,z;t)+\hat{C}(\mathbf{r},z;t)\nonumber
		\end{align}
	uniquely defines the positive-frequency field operator $\hat{A}_{\mathrm{out}}^{(+)}(\mathbf{r},z;t)$, which for $z\geq z_{\mathrm{ap}}$ describes the electromagnetic field after passing through the aperture.
	Here $\hat{C}(\mathbf{r},z;t)$ defines noise modes in the vacuum state.
	This operator is defined from the conditions of preserving the commutation relations.

    The detection system analyzes quantum states of the mode $V_{\mathrm{out}}\left(\mathbf{r},z;t\right)$ with the $P$ function
		\begin{align}\label{Eq:POut}
			P_{\mathrm{out}}(\alpha)=\frac{1}{\pi^2}\int_{\mathbb{C}}d^2 \beta \Tr\left[\hat{\rho}\,e^{(\hat{a}^\dag_{\mathrm{out}}-\alpha^\ast)\beta}e^{-(\hat{a}_{\mathrm{out}}-\alpha)\beta^\ast}\right],
		\end{align}
	where
		\begin{align}
			\hat{a}_{\mathrm{out}}=\int_{\mathbb{R}^3}d^2{\mathbf{r}}dz\hat{A}_{\mathrm{out}}^{(+)}(\mathbf{r},z;t) V_{\mathrm{out}}^{\ast}\left(\mathbf{r},z;t\right).
		\end{align}
	Using Eq.~(\ref{Eq:IOR_Position}) we can show that
		\begin{align}
			\hat{a}_{\mathrm{out}}=\sqrt{\eta}\hat{a}_{\mathrm{in}}+\sqrt{1-\eta}\hat{c},
		\end{align}
	where $\hat{c}$ is an operator representing a noise mode in a vacuum state, which is directly related to $\hat{C}(\mathbf{r},z;t)$ .
	Substituting this expression into Eq.~(\ref{Eq:POut}) and using Eq.~(\ref{Eq:PIn}) one gets
		\begin{align}
			P_{\mathrm{out}}(\alpha)=\frac{1}{\eta}P_{\mathrm{in}}\left(\frac{\alpha}{\sqrt{\eta}}\right).
		\end{align}
    The detection systems considered here do not distinguish between fluctuating shapes of the modes $V_{\mathrm{out}}\left(\mathbf{r},z;t\right)$.
	This implies that the final expression should be averaged with respect to $\eta$.
	As a result, we arrive at the input-output relation (\ref{Eq:POutDist}).

	The transmittance $\eta$ can be obtained by normalization of $V_{\mathrm{out}}\left(\mathbf{r},z;t\right)$ [cf.~Eq.~(\ref{Eq:Vout})].
	We employ the fact that $u_{\mathrm{out}}\left(\mathbf{r},z;k\right)$ varies much slower with $z$ than $e^{ikz}$.
	Together with Eq.~(\ref{Eq:Uout}) this gives
		\begin{align}
			\eta=\int_{\mathbb{R}}dk |\Phi(k)|^2 \eta(k),
		\end{align}
	where
		\begin{align}
			\eta(k)=\int_{\mathcal{A}}d^2\mathbf{r} |u_{\mathrm{in}}\left(\mathbf{r},z_{\mathrm{ap}};k\right)|^2.
		\end{align}
	Since $\Phi(k)$ is a sharp normalized function, we get
		\begin{align}
			\eta \approx\eta(k_0),
		\end{align}
	where $k_0$ is the maximum point of $|\Phi(k)|^2$.
	This leads to Eq.~(\ref{Eq:Efficiency}).
	This means that the transmittance does not depend significantly on the time spectrum $\Phi(k)$ of quasi-monochromatic modes.

\section{Verification of phase-screen method}
\label{App:Verif}

	In this appendix we present a verification technique of the sparse-spectrum model for the phase-screen method.
	For this purpose we consider the phase-structure function \cite{Tatarskii,Andrews_book}
		\begin{align}
			D_{\phi}&=\left\langle \left[ \phi(\mathbf{r} )-\phi(\mathbf{r}+\Delta\mathbf{r} ) \right]^2\right\rangle\nonumber\\
							&=2\left[\left\langle \phi^2(\mathbf{r}) \right\rangle-\left\langle \phi(\mathbf{r})\phi(\mathbf{r}+\Delta\mathbf{r}) \right\rangle\right]
		\end{align}
	where the correlation functions are given by Eq.~(\ref{Eq:Correlator_phi}).
	We compare these theoretical values with the sampled values obtained by three different methods:
	(i) the standard phase-screen method with a fast Fourier transform (FFT) \cite{Fleck1976,Frehlich2000,Lukin_book,Schmidt_book},
	(ii) the method of subharmonics \cite{Lane1992},
	and (iii) the sparse-spectrum model \cite{Charnotskii2013a,Charnotskii2013b,Charnotskii2020}.

\begin{figure}[ht!]
			\includegraphics[width=1\linewidth]{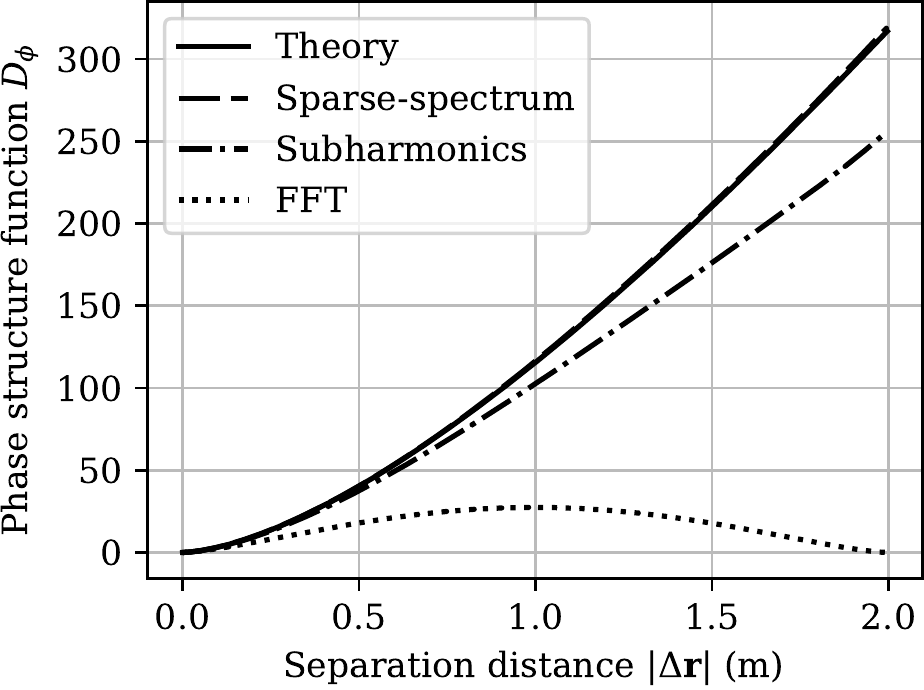}
			\caption{\label{Fig:StrFunc} Phase-structure function depending on the separation distance.
				The lines correspond to the theoretical value and to the results of three different simulations.}
		\end{figure}

	The results are shown in Fig.~\ref{Fig:StrFunc}.
	It can be clearly seen that the sparse-spectrum model gives the best agreement between theory and numerical data.
	For our simulation we use the following parameters of the atmosphere and the simulation parameters: $\lambda=808$~nm, $C_n^2=10^{-14}$~m$^{-2/3}$, $\ell_0=10^{-3}$~m, $L_0=80$~m, $l=z_m-z_{m-1}=100$~m [cf. Eq.~(\ref{Eq:Phase})], a spatial grid composed of 1024 points for each dimension, a spatial grid step of 2~mm, and the number of iterations $10^{4}$.
	For the subharmonic method, we use six subharmonics.
	For the sparse-spectrum model, the number of spectral rings is $N=1024$ and the inner and outer bounds of the spectrum are $K_\mathrm{min}=1/15 L_0$ and $K_\mathrm{max}=2/\ell_0$, respectively.

\section{Beam-wandering PDT}
\label{App:BW}
	As it has been shown in Ref.~\cite{vasylyev12}, the beam-wandering PDT is given by the log-negative Weibull distribution, i.e.,
        \begin{align}\label{Eq:BW_PDT}
            \mathcal{P}(\eta)=&\frac{R^2}{\sigma_{\mathrm{bw}}^2\eta\,\vartheta(2/W_\mathrm{ST})}\left(\ln\frac{\eta_0}{\eta}\right)^{\frac{2}{\vartheta(2/W_\mathrm{ST})}-1}\nonumber\\
            &\times
            \exp\left[-\frac{R^2}{2\sigma^2_{\mathrm{bw}}}\left(\ln\frac{\eta_0}{\eta}\right)^{\frac{2}{\vartheta(2/W_\mathrm{ST})}}\right]
        \end{align}
    for $\eta\in[0,\eta_0]$ and $\mathcal{P}(\eta)=0$ otherwise.
    Here
    	\begin{align}\label{Eq:Eta0}
    		\eta_0=1-\exp\left(-2\frac{R_\mathrm{ap}^2}{W_\mathrm{ST}^2}\right)
    	\end{align}
	is the maximum transmittance,
        \begin{align}\label{Eq:Scale}
            & R(\zeta)=R_\mathrm{ap}\\
            &\times\Bigl[\ln\Bigl(2\frac{\eta_0}{1-\exp\bigl(-R_\mathrm{ap}^2\zeta^2\bigr)}\I_0\bigl(R_\mathrm{ap}^2\zeta^2\bigr)\Bigr)\Bigr]^{-\frac{1}{\vartheta(\zeta)}}\nonumber
            \end{align}
   	is the scale parameter, and
        \begin{align}\label{Eq:Shape}
            \vartheta(\zeta)&=2R_\mathrm{ap}^2\zeta^2\frac{e^{-R_\mathrm{ap}^2\zeta^2}\I_1\bigl(R_\mathrm{ap}^2\zeta^2\bigr)}{1-\exp\bigl(-R_\mathrm{ap}^2\zeta^2\bigr)\I_0\bigl(R_\mathrm{ap}^2\zeta^2\bigr)}\nonumber\\
            &\times \left[\ln\left(2\frac{\eta_0 }{1-\exp\bigl(-R_\mathrm{ap}^2\zeta^2\bigr)\I_0\bigl(R_\mathrm{ap}^2\zeta^2\bigr)}\right)\right]^{-1}
        \end{align}
	is the shape parameter.
	In these equations, $R_\mathrm{ap}$ is the aperture radius, $W_\mathrm{ST}^2$ is the short term [cf. Eq.~(\ref{Eq:ST})], $\sigma_{\mathrm{bw}}^2$ is the beam-wandering variance [cf. Eq.~(\ref{Eq:SigmaBW})], $\zeta=2/W_{\mathrm{ST}}$, and $\I_n(x)$ is the modified Bessel function.


\section{Analytical approximation for the elliptic-beam model}
\label{App:EB}

	In order to formulate the analytical approximation used in the elliptic-beam model \cite{vasylyev16}, we first need an expression for the transmittance of the elliptical beam (\ref{Eq:EllipticIntens}) through the aperture of the radius $R_\mathrm{ap}$, as a function of the vector $\mathbf{v}=\begin{pmatrix}x_0 & y_0 & \Theta_1 & \Theta_2\end{pmatrix}$ and the angle $\phi$ between the axis $x$ and the ellipse semiaxis $W_1$.
	Here $x_0$ and $y_0$ are the beam-centroid coordinates and $\Theta_{1,2}$ are related to the length of the ellipse semiaxes, $W_{1,2}$, as given by Eq.~(\ref{Eq:Theta}).
 	This transmittance reads
	 	\begin{align}\label{Eq:etaElBeam}
	 		\eta=\widetilde\eta_0\exp\left\{-\left[\frac{r_0}{R\left(\frac{2}{W_{\rm eff}(\phi-\chi)}\right)}\right]^{\vartheta\left(\frac{2}{W_{\rm eff}(\phi-\chi)}\right)}\right\},
	 	\end{align}
 	where $r_0=|\mathbf{r}_0|$ is the beam-deflection distance; the scale parameter $R(\zeta)$ and the shape parameter $\vartheta(\zeta)$ are given by Eqs.~(\ref{Eq:Scale}) and (\ref{Eq:Shape}), respectively;
		\begin{align}
			\widetilde\eta_0&=1{-}\I_0\left(R_\mathrm{ap}^2\frac{W_1^2-W_2^2}{W_1^2W_2^2}\right)e^{-R_\mathrm{ap}^2\frac{W_1^2+W_2^2}{W_1^2W_2^2}}\nonumber\\
			&-2\Bigl[1{-}e^{-\frac{R_\mathrm{ap}^2}{2}\Bigl(\frac{1}{W_1}{-}\frac{1}{W_2}\Bigr)^2}\Bigr]\\
			&\times\exp\left[-\left\{\frac{R_\mathrm{ap}\frac{(W_1+W_2)^2}{|W_1^2-W_2^2|}}{R\left(\frac{1}{W_1}-\frac{1}{W_2}\right)}\right\}^{\vartheta\left(\frac{1}{W_1}-\frac{1}{W_2}\right)}\right],\nonumber
		\end{align}
	is the maximum transmittance for the elliptic beam;
		\begin{align}
			& W_{\rm eff}^2(\phi-\chi)=4R_\mathrm{ap}^2\Bigl[\mathcal{W}\Bigl(\frac{4R_\mathrm{ap}^2}{W_1W_2}e^{2R_\mathrm{ap}^2\left(\frac{1}{W_1^2}+\frac{1}{W_2^2}\right)}\Bigr.\Bigr.\\
			&\qquad\times\Bigl.\Bigl. e^{R_\mathrm{ap}^2\left(\frac{1}{W_1^2}-\frac{1}{W_2^2}\right)\cos(2\phi-2\chi)}\Bigr)\Bigr]^{-1};\nonumber
		\end{align}
	$\chi$ is the angle that defines the direction of the vector $\mathbf{r}_0$ to the $x$ axis (cf. Fig.~\ref{Fig:img_corr_r0_wr}); $\I_n(x)$ is the modified Bessel function; and $\mathcal{W}$ is the Lambert  function.

	The elliptic-beam PDT is given by
		\begin{align}\label{Eq:PDTC_EB}
			\mathcal{P}\left(\eta\right){=}\frac{2}{\pi}\int_{\mathbb{R}^4}d^4\mathbf{v}
			\int\limits_{0}^{ \pi /2}d\phi \,
			\rho_G(\mathbf{v}|\boldsymbol{\mu},\Sigma)\delta\left[\eta{-}\eta\left(\mathbf{v},\phi\right)\right].
		\end{align}
	Here $\eta\left(\mathbf{v},\phi\right)$ is the transmittance defined by Eq.~(\ref{Eq:etaElBeam}) as a function of random parameters and
	$\rho_G(\mathbf{v}|\boldsymbol{\mu},\Sigma)$ is the Gaussian probability density function of the vector $\mathbf{v}$ with the mean  $\boldsymbol{\mu}$ and the covariance matrix $\Sigma$.
	This matrix is clearly divided into two diagonal blocks since $\langle\Delta\mathbf{r}_0\Delta\Theta_{1,2}\rangle=0$ due to symmetry.
	The first block, related to the beam-centroid coordinates $x_0$ and $y_0$, reads $\sigma_\mathrm{bw}^2\mathbb{I}_2$, where $\mathbb{I}_2$ is the $2\times2$ unity matrix.
	The second block consists of the elements
		\begin{align}\label{Eq:ThetaCovariances}
			\langle \Delta\Theta_i\Delta\Theta_j\rangle=
			\ln\left[1+
			\frac{\langle \Delta W_i^2 \Delta W_j^2\rangle}{\langle
				W_i^2\rangle\langle
				W_j^2\rangle}\right].
		\end{align}
	Two first components of the vector $\boldsymbol{\mu}$ are zero in the case when $\langle\mathbf{r}_0\rangle=0$.
	The two other components are given by
		\begin{align}\label{Eq:ThetaMean}
			\langle
			\Theta_{i}\rangle=\ln\left[\frac{\langle
				W_{i}^2\rangle}{W_0^2}\left(1+
			\frac{\langle (\Delta	W_{i}^2)^2\rangle}{\langle
				W_{i}^2\rangle^2}\right)^{-1/2}\right].
		\end{align}
	The moments of $W_{i}^2$ in Eqs.~(\ref{Eq:ThetaCovariances}) and (\ref{Eq:ThetaMean}) are expressed in terms of field correlation functions $\Gamma_2\!\left(\mathbf{r};z_\mathrm{ap}\right)$ and $\Gamma_4\!\left(\mathbf{r}_1,\mathbf{r}_2;z_\mathrm{ap}\right)$ as
		\begin{align}\label{Eq:W12SqMean}
			\langle&
			W_{i}^2\rangle{=}4\left[\int_{\mathbb{R}^2}d^2\mathbf{r}\, x^2 \Gamma_2\!
			\left(\mathbf{r};z_\mathrm{ap}\right){-}
			\langle x_0^2\rangle\right],
		\end{align}
		\begin{align}
			&\langle
			W_{i}^2W_{j}^2\rangle=8\Big[{-}8\,\delta_{ij}\langle
			x_0^2\rangle^2
			{-}\langle x_0^2\rangle \langle
			W_{i}^2\rangle
			\label{Eq:W12ViaGamma}\\
			&+\int_{\mathbb{R}^4}d^4\mathbf{r}
			\,
			\left[x_1^2x_2^2\left(4\delta_{ij}{-}1\right)-x_1^2y_2^2\left(4\delta_{ij}{-}3\right)\right]\nonumber\\
			&\times
			\,\Gamma_4\!\left(\mathbf{r}_1,\mathbf{r}_2;z_\mathrm{ap}\right)
			\Big].\nonumber
		\end{align}
	These moments can be calculated analytically, for instance, using the phase approximation of the Huygens-Kirchhoff method, as demonstrated in Ref.~\cite{vasylyev16}. However, in this paper we calculated these moments from the numerically simulated data.


\section{Approximation of weak beam-wandering for the model based on the law of total probability}
\label{App:TPL}

	In both variants of the model based on the law of total probability, i.e., those based on the truncated log-normal distribution (cf. Sec.~\ref{Sec:LTP}) and those based on the Beta distribution (cf. Sec.~\ref{Sec:BetaModel}), one needs to evaluate the conditional moments of the transmittance, $\langle\eta\rangle_{r_0}$ and $\langle\eta^2\rangle_{r_0}$.
	In the approximation of weak beam wandering \cite{vasylyev18} these quantities are evaluated as
		\begin{align}\label{Eq:CondEta1}
			\langle\eta\rangle_{r_0}=\eta_0\exp\left[-\left(\frac{r_0}{R(2/W_\mathrm{ST})}\right)^{\vartheta(2/W_\mathrm{ST})}\right],
		\end{align}
		\begin{align}\label{Eq:CondEta2}
			\langle\eta^2\rangle_{r_0}=h_0^2\exp\left[-2\left(\frac{r_0}{R(2/W_\mathrm{ST})}\right)^{\vartheta(2/W_\mathrm{ST})}\right],
		\end{align}
	where the scale parameter $R(\zeta)$ and the shape parameter $\vartheta(\zeta)$ are given by Eqs.~(\ref{Eq:Scale}) and (\ref{Eq:Shape}), respectively.
	The efficiency $\eta_0$ can be obtained from Eq.~(\ref{Eq:Eta0}) if the field correlation function $\Gamma_2(\mathbf{r};z_\mathrm{ap})$ is Gaussian.
	Otherwise, it is obtained as
		\begin{align}\label{Eq:Eta1Defin}
			\eta_0=\frac{\langle\eta\rangle}{\displaystyle{\int\limits_{0}^\infty  d u\,u 	\,e^{-\frac{u^2}{2}}e^{-\left(\frac{\sigma_{\mathrm{bw}}}{R(2/W_\mathrm{ST})}u\right)^{\vartheta(2/W_\mathrm{ST})}}}}.
		\end{align}
	The parameter $h_0$ is evaluated as
		\begin{align}\label{Eq:Eta2Defin}
			h_0^2=\frac{\langle\eta^2\rangle}{\displaystyle{\int\limits_{0}^\infty  d u\,u 	\,e^{-\frac{u^2}{2}}e^{-2\left(\frac{\sigma_{\mathrm{bw}}}{R(2/W_\mathrm{ST})}u\right)^{\vartheta(2/W_\mathrm{ST})}}}}.
		\end{align}
	In addition, this approximation ensures that two important statistical moments of transmittance, namely, $\langle\eta\rangle$ and $\langle\eta^2\rangle$, retain their original definitions given by Eqs.~(\ref{Eq:Eta}) and (\ref{Eq:Eta2}), respectively.

\bibliography{biblio}
\end{document}